\documentclass{PoS}

\usepackage{caption}
\usepackage{subcaption}
\usepackage[inline]{enumitem}
\usepackage{lineno}

\title{The upgraded Pixel detector and the commissioning of the Inner Detector tracking of the ATLAS experiment for Run-2 at the Large Hadron Collider}

\ShortTitle{The upgraded Pixel detector and ID tracking of the ATLAS experiment for Run-2}

\author{\speaker{Karolos Potamianos}, on behalf of the ATLAS Collaboration\\
        Lawrence Berkeley National Laboratory\\
        E-mail: \email{karolos.potamianos@cern.ch}}

\abstract{Run-2 of the Large Hadron Collider (LHC) will provide new challenges to track and vertex reconstruction with higher energies, denser jets and higher rates. Therefore the ATLAS experiment has constructed the first 4-layer Pixel Detector in HEP, installing a new pixel layer, also called Insertable B-Layer (IBL). The IBL is a fourth layer of pixel detectors, and has been installed in May 2014 at a radius of 3.3 cm between the existing Pixel Detector and a new smaller radius beam-pipe. The new detector, built to cope with the high radiation and expected occupancy, is the first large scale application of 3D sensors and CMOS 130~nm readout electronics. In addition, the Pixel Detector was improved with a new service quarter panel to recover about 3\% of defective modules lost during Run-1 and a new optical readout system to readout the data at higher speed while reducing the occupancy when running with increased luminosity.

Complementing detector improvements, many improvements to Inner Detector track and vertex reconstruction were developed during the two-year shutdown of the LHC. These include novel techniques developed to improve the performance in the dense cores of jets, optimisation for the expected conditions, and a software campaign which lead to a factor of three decrease in the CPU time needed to process each recorded event.}

\FullConference{The European Physical Society Conference on High Energy Physics\\
		22--29 July 2015\\
		Vienna, Austria}

\begin{document}

\section{Introduction}

During Run-1 of the Large Hadron Collider (LHC), the ATLAS experiment~\cite{ref:AtlasExperiment} successfully recorded about $27$~fb$^{-1}$ of $pp$ collision data at center-of-mass energies of $\sqrt{s} =  7$ and $8$~TeV and with a bunch crossing spacing of $50$~ns.

During the Long Shutdown 1 (LS1, 2013-2014), the ATLAS experiment was opened up for maintenance and upgrades. In particular, the Pixel Detector was brought to the surface to be re-fitted with new service quarter panels (nSQP) and new optical connections. During this time, a new detector, the Diamond Beam Monitor (DBM) was inserted within the Pixel Detector volume. Finally, the Pixel Detector was extended with the first 4$^{\rm th}$ pixel layer in HEP, the Insertable-B-Layer (IBL)~\cite{ref:IBLTDR}. Also during LS1, as part of a general software overhaul, the track reconstruction algorithms were significantly improved compared to Run-1.

In 2014-2015, the improved Pixel Detector was commissioned in preparation for Run-2, which started on June 3, 2015.
During Run-2, the LHC is expected to deliver collisions at the design luminosity of $1\times 10^{34}$~cm$^{-2}$~s$^{-1}$ with a bunch crossing spacing of $25$~ns. The ATLAS experiment is set to collect over $100$~fb$^{-1}$ of $pp$ collision data.

\begin{figure}
\begin{subfigure}{.6\linewidth}
\includegraphics[width=\linewidth]{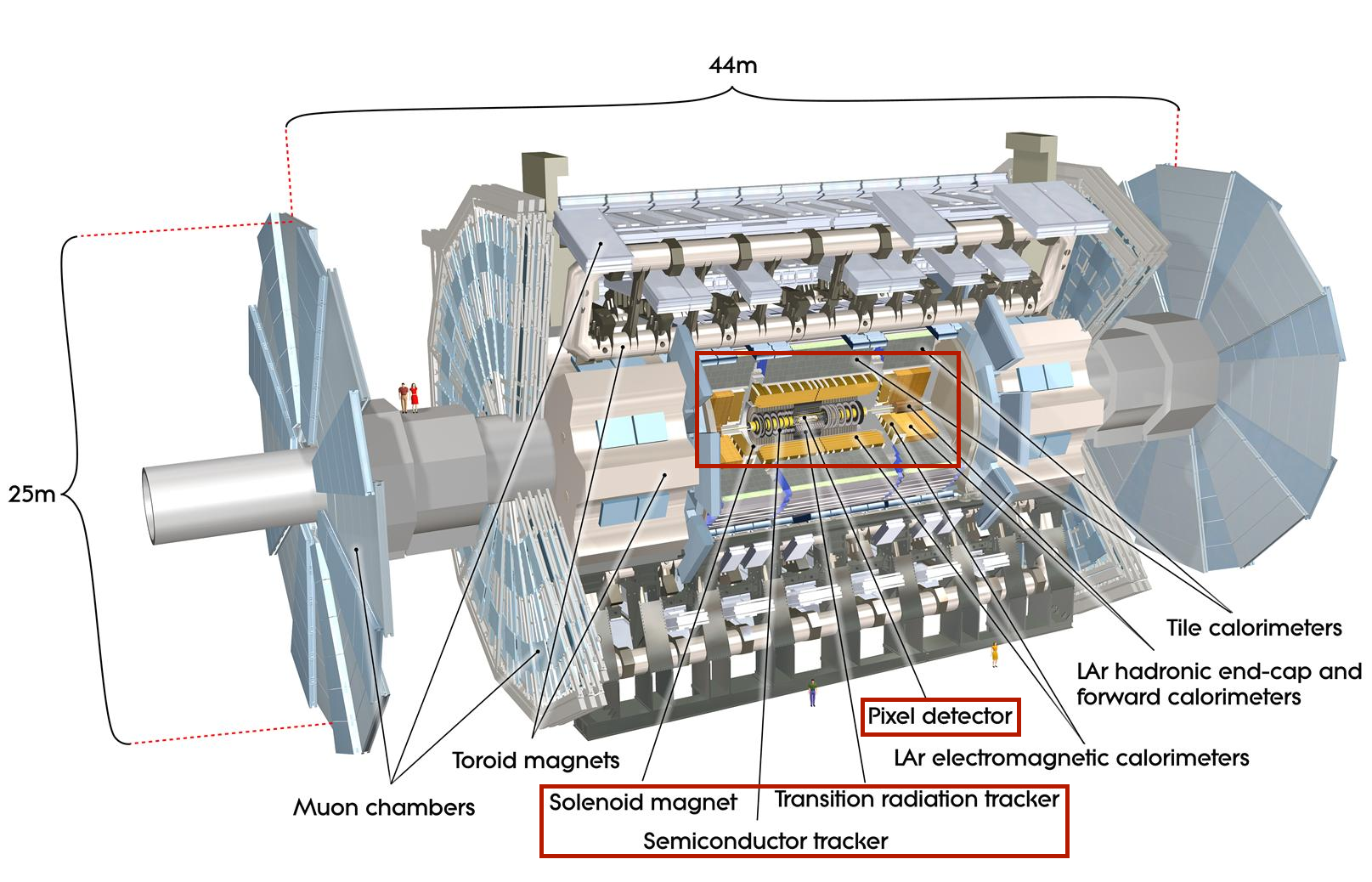}
\caption{\label{fig:AtlasDetector:a}}
\end{subfigure}
\hspace*{\fill} %
\begin{subfigure}{.4\linewidth}
\includegraphics[width=\linewidth]{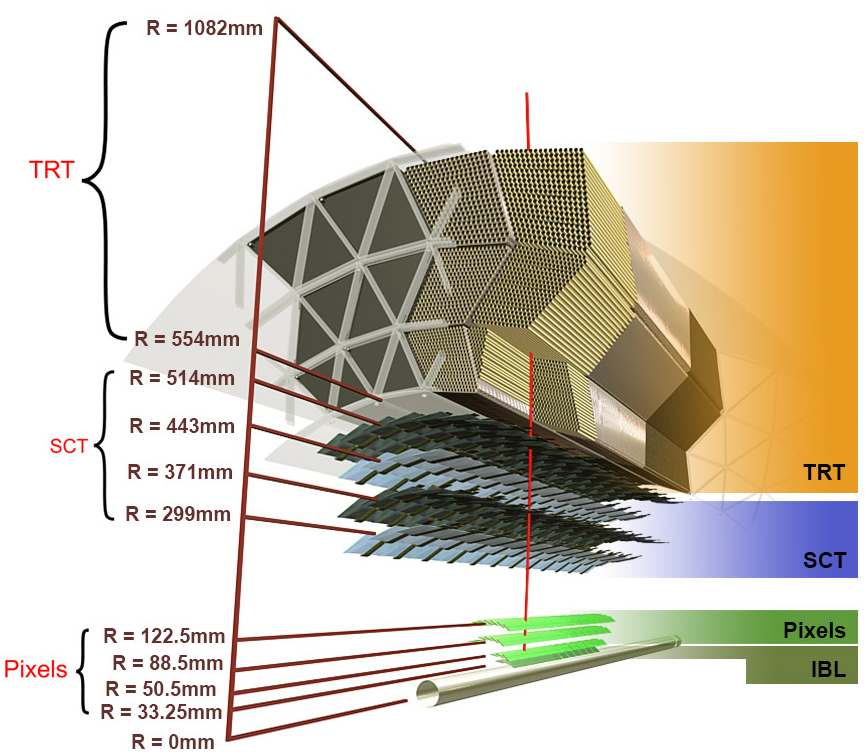}
\caption{\label{fig:AtlasDetector:b}}
\end{subfigure}
\caption{\label{fig:AtlasDetector} Schematic view of (a) the ATLAS detector, with (b) a detailed layout of the Inner Detector (ID), including the new Insertable B-Layer (IBL).}
\end{figure}

The ATLAS detector~\cite{ref:AtlasExperiment}, shown in fig.~\ref{fig:AtlasDetector:a}, is a general-purpose particle detector operated at the LHC. Its Inner Detector (ID) measures the properties of charged particles and reconstructs their trajectories (tracking) up to a pseudorapidity\footnote{ATLAS uses a right-handed coordinate system with its origin at the nominal interaction point (IP) in the center of the detector and the $z$-axis along the beam pipe. The $x$-axis points from the IP to the center of the LHC ring, and the $y$-axis points upwards. Cylindrical coordinates $(R,\phi)$ are used in the transverse plane, $\phi$ being the azimuthal angle around the $z$-axis. The pseudorapidty is defined in terms of the polar angle $\theta$ as $\eta = - \ln \tan(\theta /2)$. Angular distance is measured in units of $\Delta R \equiv \sqrt{(\Delta\eta)^2 +(\Delta\phi)^2}$.} of $|\eta|<2.5$. The ID, whose radial layout is shown in fig.~\ref{fig:AtlasDetector:b}, includes three sub-systems: the Pixel Detector (including the IBL), the Silicon strip tracker (SCT), and the transition radiation tracker (TRT).

\begin{figure}
\begin{subfigure}{.32\linewidth}
\includegraphics[width=\linewidth]{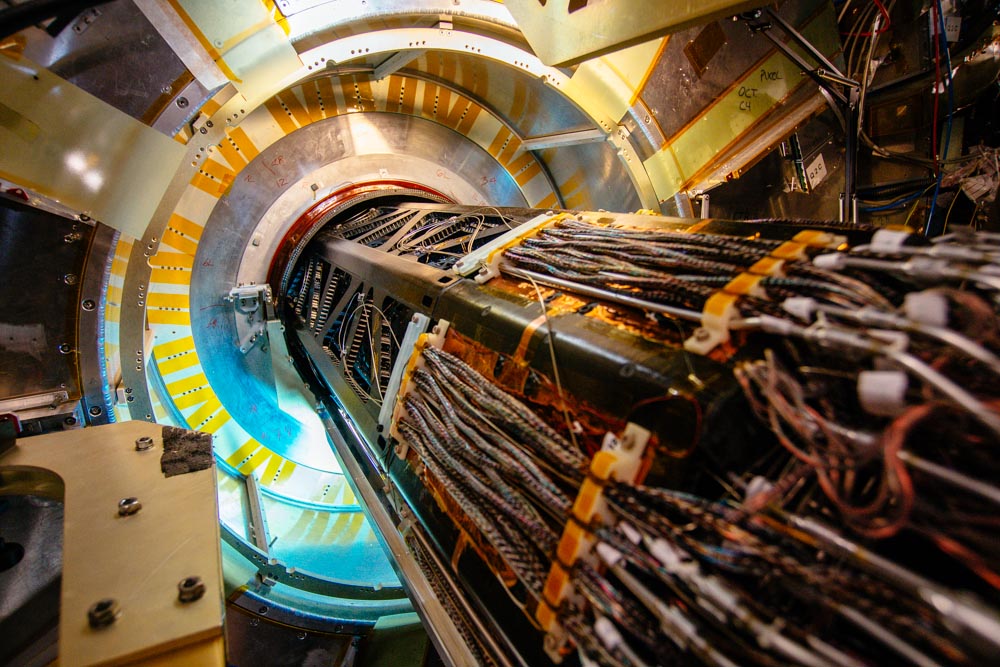}
\caption{\label{fig:PixelPictures:a}}
\end{subfigure}
\hspace*{\fill} %
\begin{subfigure}{.32\linewidth}
\includegraphics[width=\linewidth]{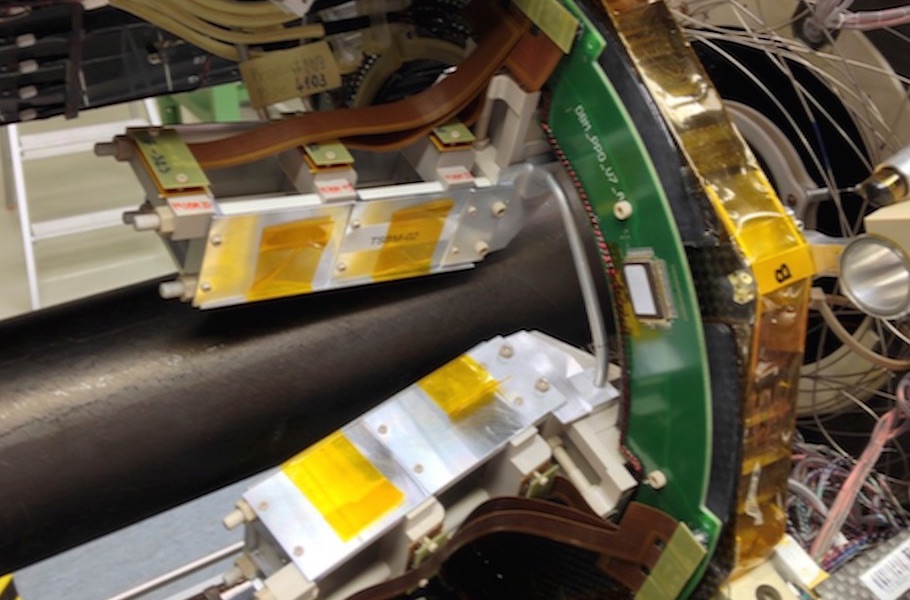}
\caption{\label{fig:PixelPictures:b}}
\end{subfigure}
\hspace*{\fill} %
\begin{subfigure}{.32\linewidth}
\includegraphics[width=\linewidth]{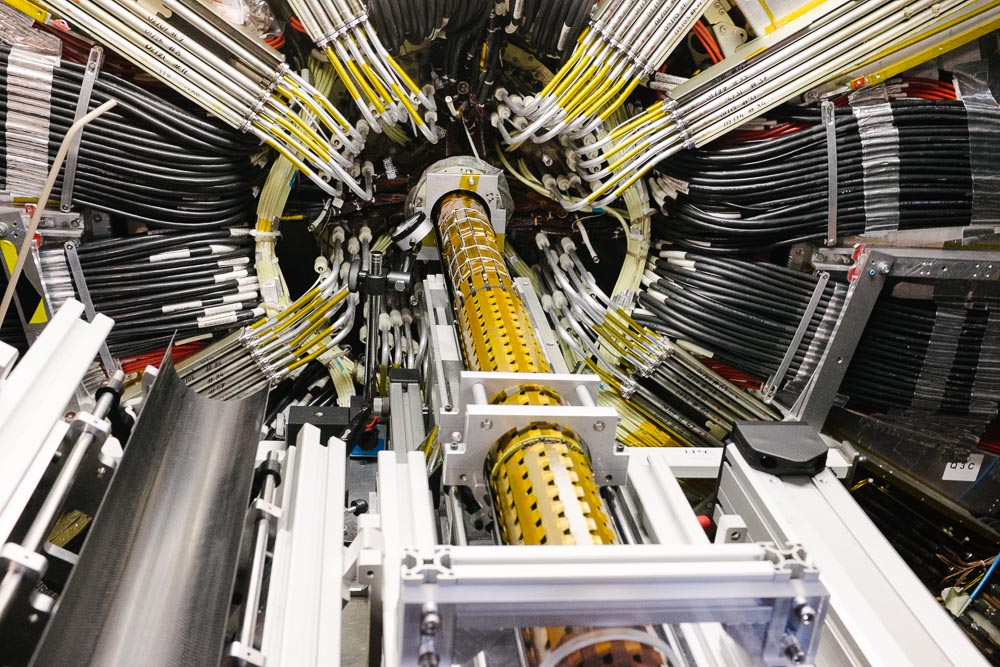}
\caption{\label{fig:PixelPictures:c}}
\end{subfigure}
\caption{\label{fig:PixelPictures}Upgrades to the ATLAS Pixel Detector during the Long Shutdown 1 (LS1): (a) the Pixel Detector with new services and new optical links ; (b) the Diamond Beam Monitor (DBM) inside the Pixel Detector volume ; and (c) the new Insertable B-Layer (IBL), the 4$^{\rm th}$ Pixel Detector layer of ATLAS.}
\end{figure}

Figure~\ref{fig:PixelPictures} shows the various steps of the LS1 Pixel Detector upgrade program. The upgrade of the Pixel Detector and the IBL are presented in Section~\ref{sec:IBL}. In Section~\ref{sec:IDTrackingAndVertexReco}, we review the major improvements to track and vertex reconstruction as well as improvements in $b$-jet identification. Performance measurements using commissioning and early Run-2 data are presented in Section~\ref{sec:Commissioning}.

\section{\label{sec:IBL}The upgraded Pixel Detector and the new Insertable B-Layer (IBL)}

The Run-1 ATLAS Pixel Detector~\cite{ref:AtlasPixel} consists of 3 barrel layers located at radii of $50.5$ (B-Layer), $88.5$ (Layer 1), and $122.5$~mm (Layer 2) centered around the beam axis, and 2 end-caps with 3 disc layers each. In total, 1744 modules read out 80 million pixels. The typical pixel size is $50~\mu{\rm m} \times 400~\mu{\rm m}$ and the sensor has a thickness of $250~\mu m$. 

\begin{figure}
\centering
\includegraphics[width=.5\linewidth]{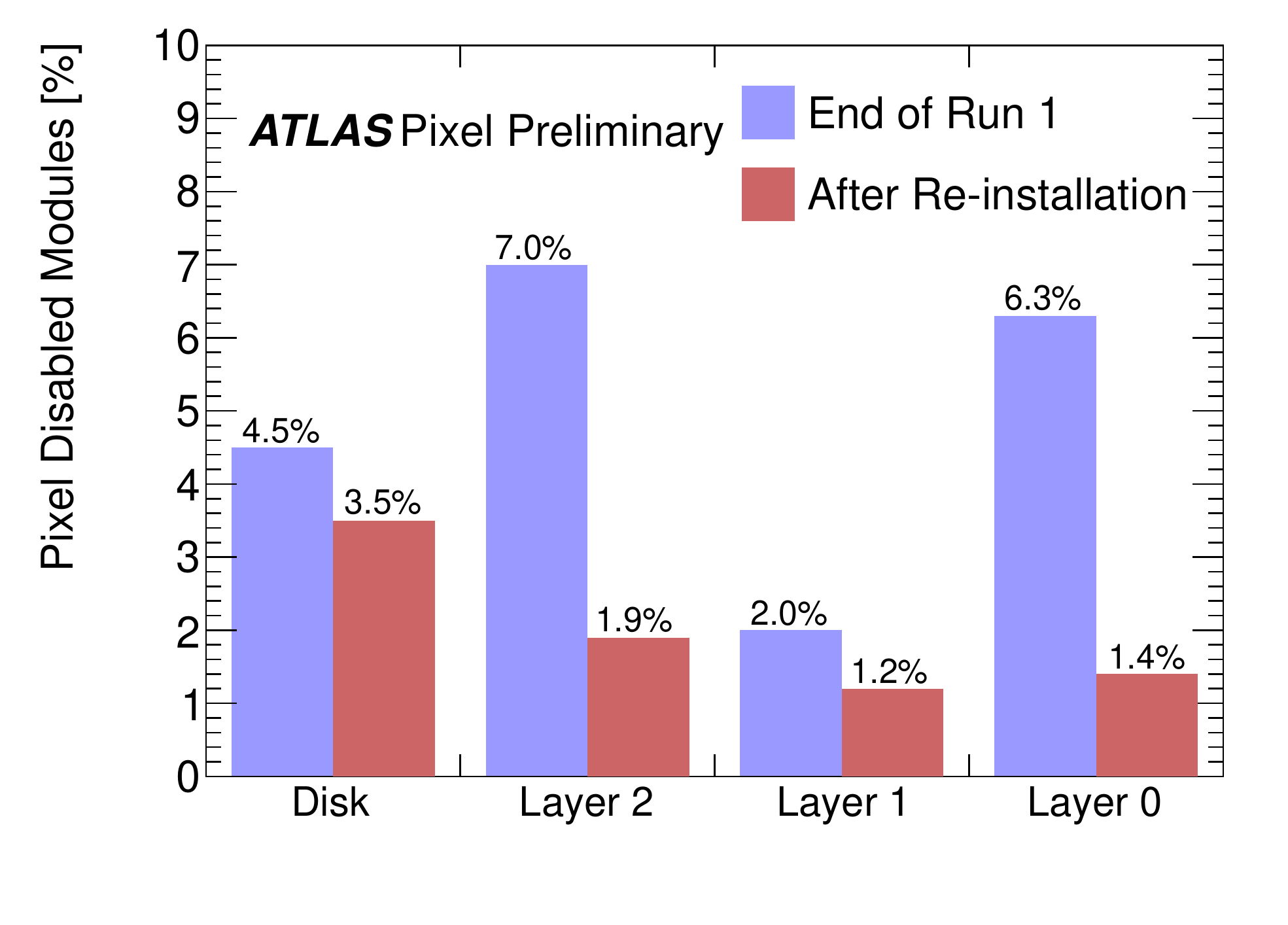}
\caption{\label{fig:PixDisabledByLayer}Fraction of disabled modules in the Run-1 Pixel Detector before and after re-installation in the ATLAS experimental cavern, following an update of the services (nSQP) and installation of new optical links, and module repairs~\cite{ref:PixelPublicPlots}.}
\end{figure}

To transfer the data out of the Pixel Detector, optoboards are used to convert electrical signals into optical ones. In Run-1 these boards were located inside the detector volume and were not accessible for repair. In order to make them accessible for repairs, the Pixel Detector was extracted and brought to the surface to have its services replaced by the nSQPs, which extend the electrical lines outside the detector volume, and allow for installation of the optoboards outside it. In addition, additional optical fibres were installed for Layer 1 to increase the data transmission rate from $80$ to $160$~Mbps which reduces the number of desynchronized modules caused by high hit occupancy in the detector modules. These additional fibres will be used in the upcoming readout upgrade foreseen for the end of 2015 (Layer 2) and that of 2016 (Layer 1). The effect of the installation of the nSQPs and that of various repairs can be seen in fig.~\ref{fig:PixDisabledByLayer}~\cite{ref:PixelPublicPlots}.

Designed for an instantaneous luminosity of up to $1 \times 10^{34}~{\rm cm}^{-2}{\rm s}^{-1}$, the Run-1 Pixel Detector will be under heavy load with the higher luminosities expected during and after Run-2, the limiting factors being the effects of possible radiation damage, the data transmission speed and the module buffer size. This motivates the nSQP and readout upgrades.

\begin{figure}
\begin{subfigure}{.45\linewidth}
\includegraphics[width=\linewidth]{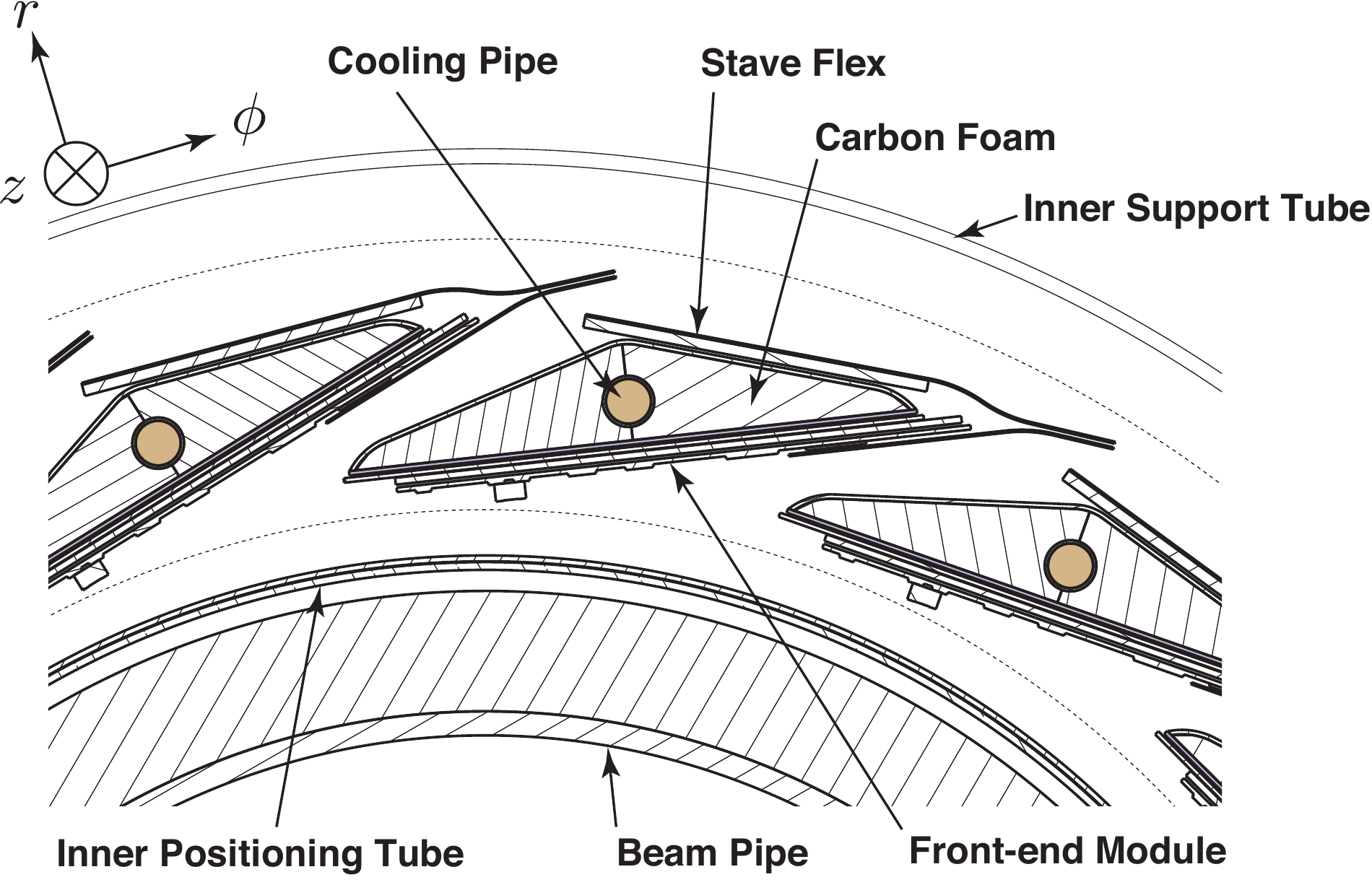}
\caption{\label{fig:IBL:a}}
\end{subfigure}
\begin{subfigure}{.55\linewidth}
\includegraphics[width=\linewidth]{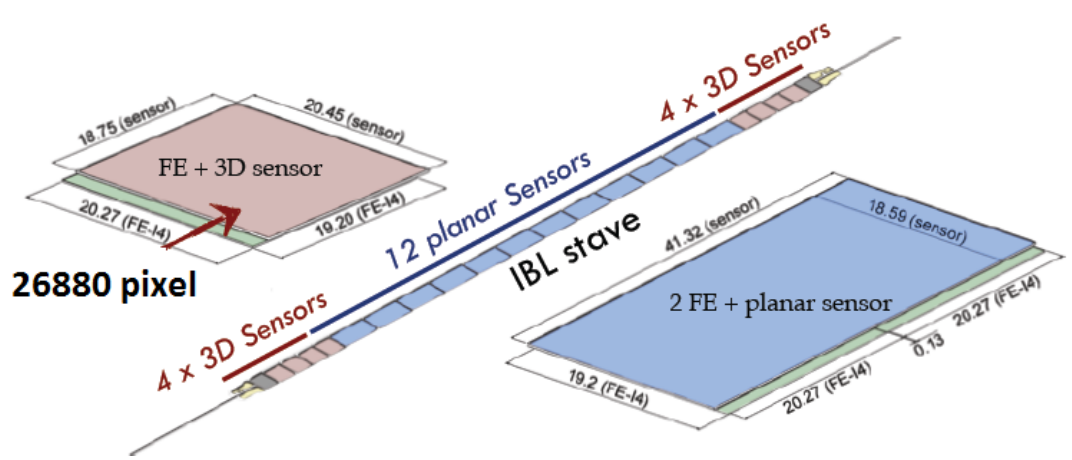}
\caption{\label{fig:IBL:b}}
\end{subfigure}
\caption{\label{fig:IBL} (a) Transverse view of 3 of the Insertable-B-Layer (IBL) staves, located directly on the beam pipe. (b) The layout of one of the 14 IBL staves, comprising six dual-chip modules with planar sensors (in the central part) and two groups of four  single chip modules with 3D sensors (at the stave edges).
Each front-end chip has a size of approximately 4~cm$^2$ and contains 26880 pixels. }
\end{figure}

In addition to these improvements, the Pixel Detector was expanded by inserting a new, innermost layer: the IBL~\cite{ref:IBLTDR}, located at a mean radius of $33.2$~mm around the beam pipe (of smaller radius than in Run-1), and containing 12 million pixels with a typical size of $50~\mu{\rm m} \times 250~\mu{\rm m}$. Figure~\ref{fig:IBL} shows the IBL within the Pixel Detector volume and around the beam pipe. It is the first large scale application of 3D detectors and CMOS $130$~nm chip readout.
The IBL consists of 14 staves with each 20 modules made using either planar or 3D sensors (with a thickness of $200$ and $230$~$\mu$m respectively) with 2 or 1 FE-I4 front-end bump-bonded chips, respectively.
The temperature of the IBL is controlled using a bi-phase CO$_2$ cooling system. Details on the IBL and its quality assurance can be found in~\cite{ref:iblStaveQA}.

\begin{figure}
\centering
\includegraphics[width=.5\linewidth]{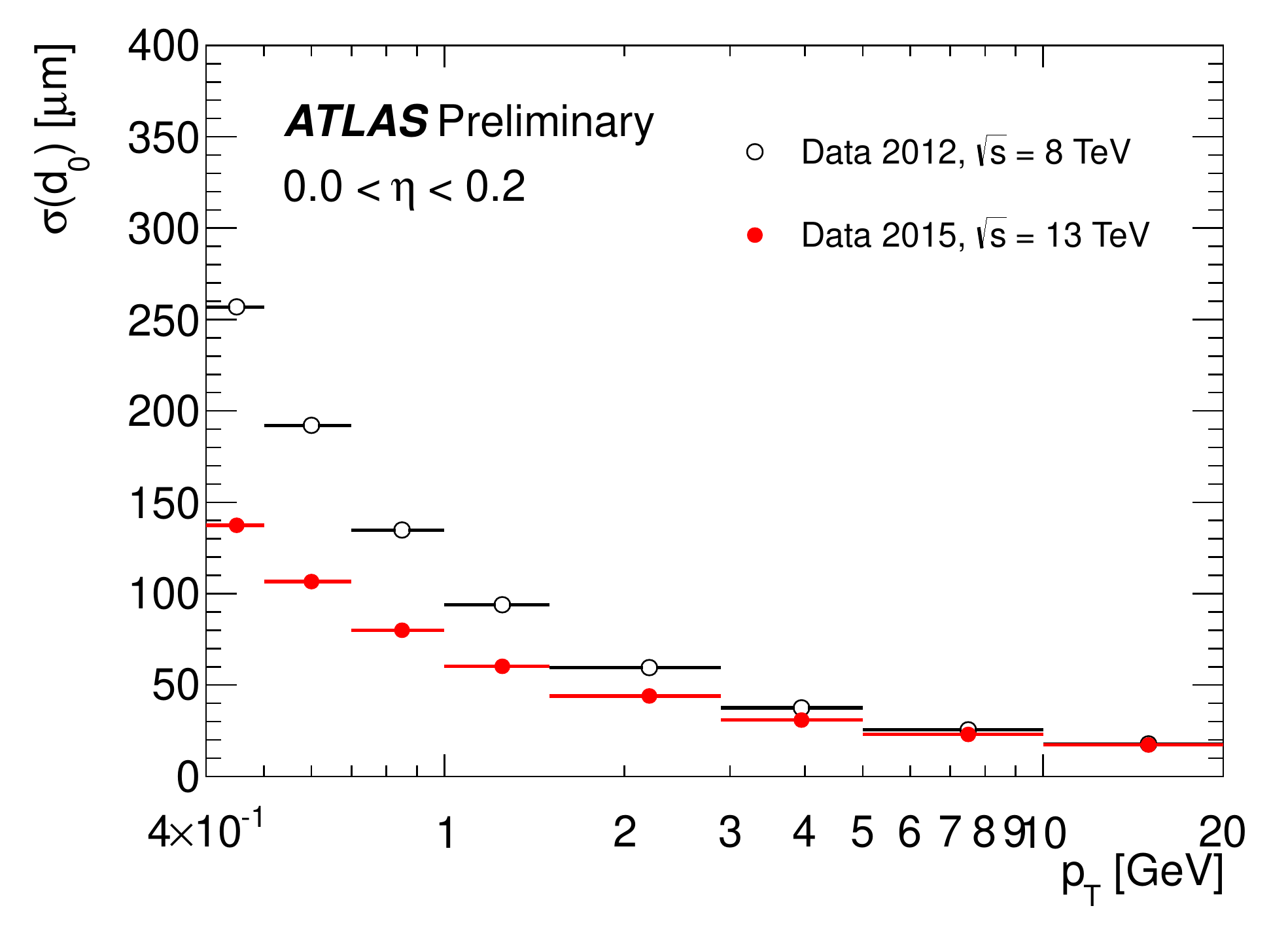}
\caption{\label{fig:TransverseImpactParameterRes} Unfolded transverse impact parameter resolution measured from data in 2015, at $\sqrt{s}~=~13$~TeV, with the Inner Detector including the IBL, as a function of track $p_T$, for values of $0.0 < \eta < 0.2$, compared to that measured from data in 2012, $\sqrt{s} = 8$~TeV. The data in 2015 is collected with a minimum bias trigger. The data in 2012 is derived from a mixture of jet, tau and missing $E_T$ triggers~\cite{ref:IDTR-2015-007}.}
\end{figure}

The IBL improves tracking by providing an additional measurement point, and mitigates the possible loss of hits in the 3 layers foreseen with higher instantaneous luminosity, and after radiation damage. Figure~\ref{fig:TransverseImpactParameterRes} shows the improvement in impact parameter resolution due to the IBL as measured from early Run-2 data with respect to Run-1.

Run-2 and Run-3 of the LHC are expected to deliver around $300$~fb$^{-1}$ of integrated luminosity. The Run-1 luminosity monitor of ATLAS, the Beam Conditions Monitor (BCM) will saturate around an instantaneous luminosity of $1 \times 10^{34}~{\rm cm}^{-2}{\rm s}^{-1}$. Therefore, a new detector, the Diamond Beam Monitor (DBM) was inserted inside of the Pixel Detector volume. It adds 280 000 channels, which allow the DBM to exploit the space configuration of hits and tracks. The DBM has 4 tracking telescopes (3 with diamond and 1 with Silicon sensors) consisting of 3 layers of sensors with FE-I4 chips on either side of the interaction point, at $|\eta| \sim 3.2$. The advantage of diamond over Silicon is that it is much more radiation tolerant at all energies, which is important for the precision and stability of luminosity measurements as the detector's exposure to radiation increases.

\section{\label{sec:IDTrackingAndVertexReco}Improvements to Inner Detector track and vertex reconstruction}

\subsection{Improvements to the tracking software}

\begin{figure}
\centering
\includegraphics[width=.5\linewidth]{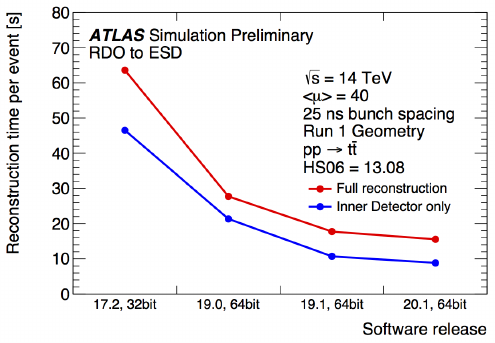}
\caption{\label{fig:TrackRecoTime}Total and ID-only reconstruction time per event while reconstructing the same simulated sample using the last Run-1 software release up to early Run-2 releases~\cite{ref:OfflineRecoSwTiming}.}
\end{figure}

For Run-2, ATLAS targets a $1$~kHz event reconstruction rate. This required significantly reducing the ID track reconstruction time, which accounts for 70\% of the total reconstruction time. A software optimization was performed during LS1, in parallel with the detector upgrades~\cite{ref:TrackRecoTiming}. Improvements include:
\begin{enumerate*}[label=(\alph*)]
\item an optimized track extrapolation code and new C++ based magnetic field description
\item a migration of the linear algebra solver from CLHEP to EIGEN
\item the use of a simplified Event Data Model (EDM), called xAOD, with track seeding optimized for high pile-up
\item the restructuring of the code to reduce inheritance and employ a template class design
\item the migration to a 64-bit architecture
\end{enumerate*}.
Figure~\ref{fig:TrackRecoTime} shows that both ATLAS and ID reconstruction times have been reduced by a factor of 4~\cite{ref:OfflineRecoSwTiming}.%

\subsection{Performance improvements: tracking in dense environments (TIDE)}

Dense environments are characterized by charged particles separated by distances of the order of the granularity of the ATLAS ID. With the IBL providing additional measurement points closer to the interaction point and an increased centre-of-mass energy, the performance of tracking in dense environments (TIDE)~\cite{ref:TIDE} will be  particularly crucial. Therefore, algorithmic optimizations were carried out in the pattern recognition stage of the track reconstruction for improved performance.

In Run-1, a neural network (NN) was used to identify, using the charge and topology of the cluster, whether a cluster is created by a single charged particle, or by several charged particles sufficiently close that their clusters are merged~\cite{ref:ClusteringNN}.
The Run-1 algorithm decided whether a cluster is allowed to be shared by multiple tracks before track seeds were identified. The proper identification of merged clusters improves the cluster assignment and thus the track reconstruction efficiency in dense environments. It also decreases the improper assignment of a cluster to multiple tracks.
In Run-2, the decision as to whether a cluster is allowed to be shared or not is delayed to the ambiguity resolution stage. At that stage, the NN is used only when multiple tracks share the same cluster, and a penalty system is introduced: if a cluster is considered shareable, the tracks sharing that cluster are not penalized, whereas they are penalized if the cluster is not shareable. This effectively reduces the number of clusters shared between tracks.

\begin{figure}
\begin{subfigure}{.5\linewidth}
\includegraphics[width=\linewidth]{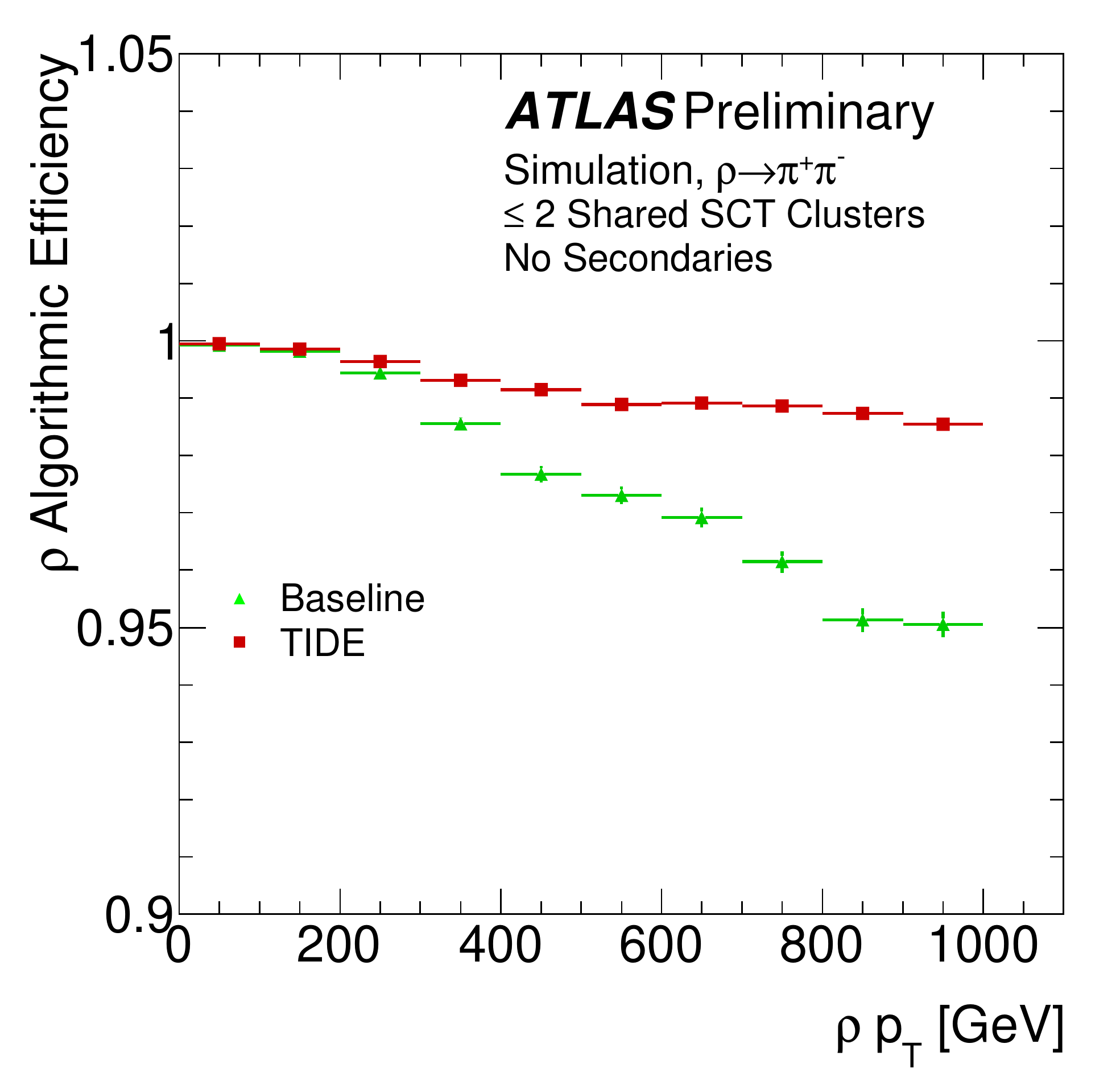}
\caption{\label{fig:DecayProductsEfficiency:a}}
\end{subfigure}
\hspace*{\fill} %
\begin{subfigure}{.5\linewidth}
\includegraphics[width=\linewidth]{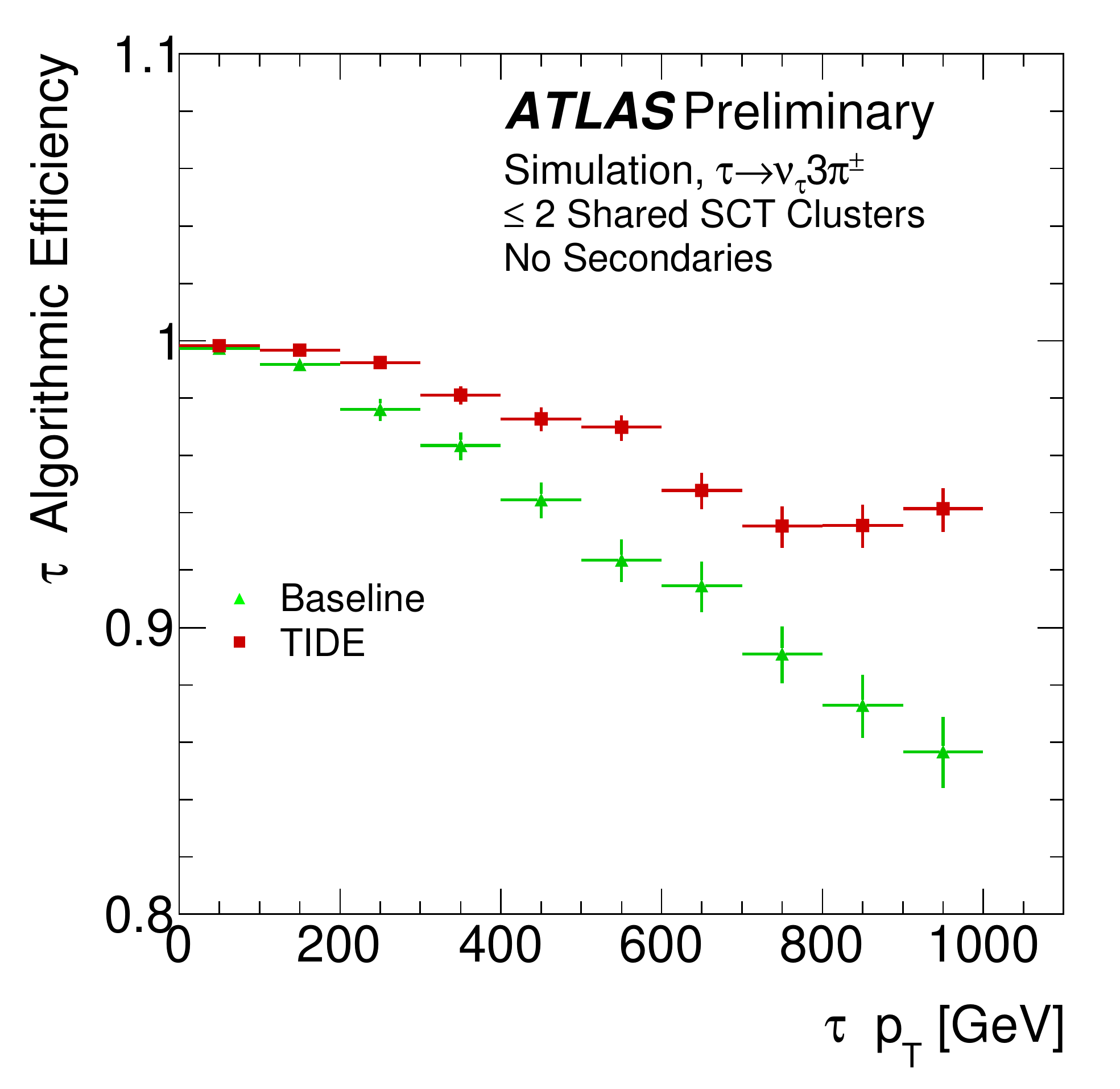}
\caption{\label{fig:DecayProductsEfficiency:b}}
\end{subfigure}
\caption{\label{fig:DecayProductsEfficiency}  The efficiency for the reconstruction of all decay products of a $\rho$ (a) or 3-prong $\tau$ (b) in events where truth-based tracks do not share more than two clusters in the SCT is shown as a function of the parent truth particle $p_T$.
The events are restricted to contain no secondary particles through nuclear interactions between the decay products and the detector material.
The performance with the new (TIDE) reconstruction (red squares) is compared with that using the baseline reconstruction from Run-1 (green triangles)~\cite{ref:TIDE}. }
\end{figure}

\begin{figure}
\begin{subfigure}{.5\linewidth}
\includegraphics[width=\linewidth]{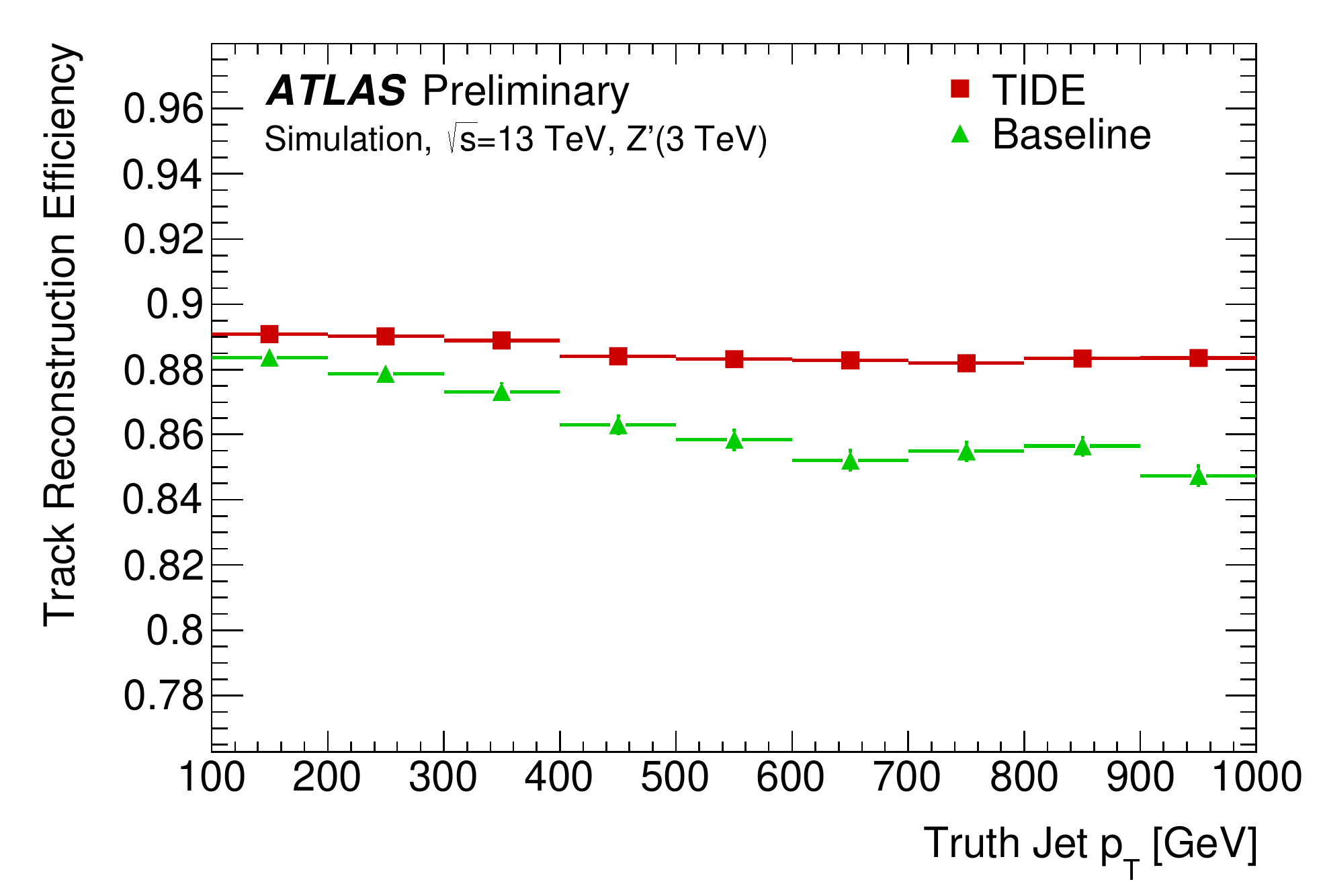}
\caption{\label{fig:TrackRecoPerf:a}}
\end{subfigure}
\hspace*{\fill} %
\begin{subfigure}{.5\linewidth}
\includegraphics[width=\linewidth]{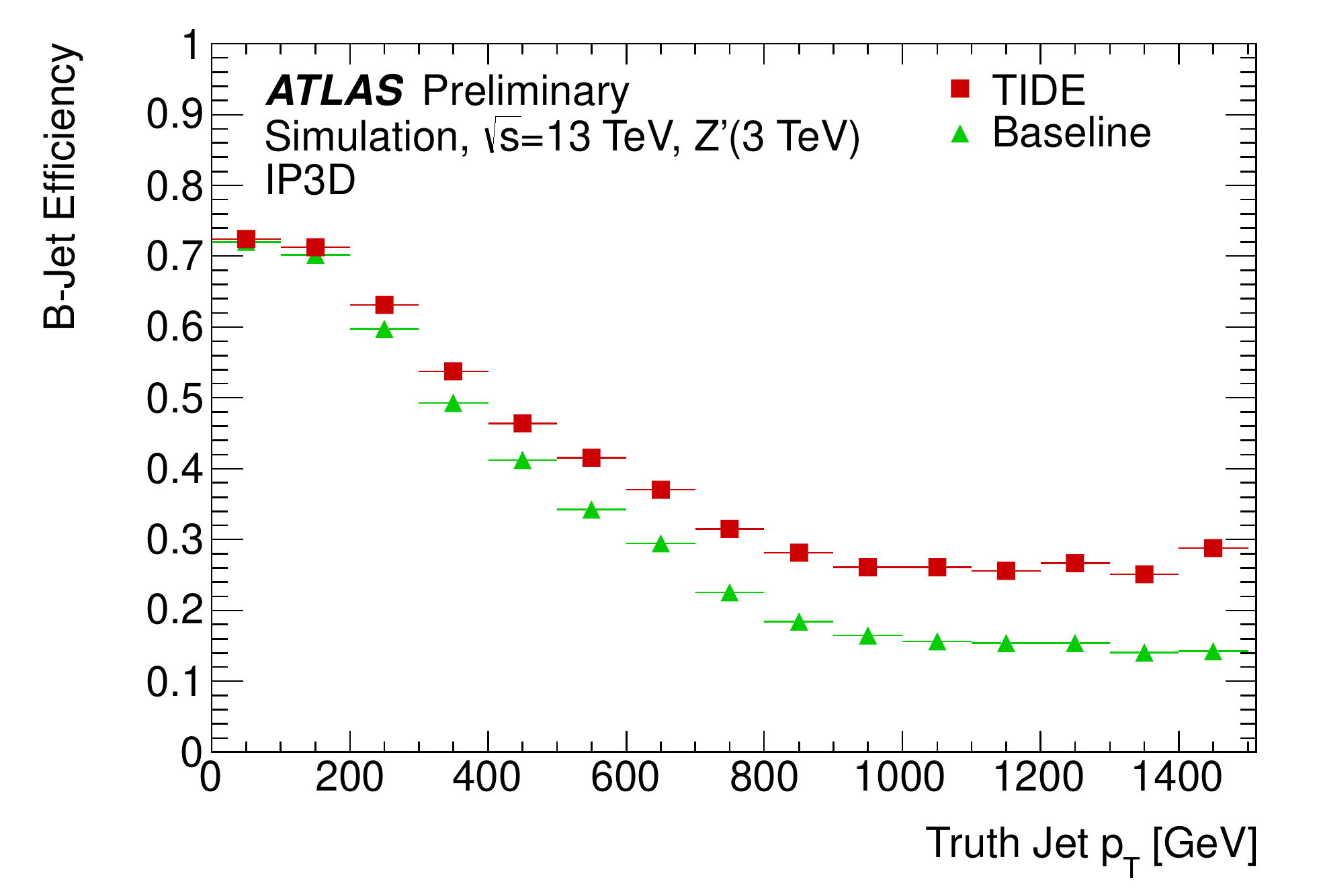}
\caption{\label{fig:TrackRecoPerf:b}}
\end{subfigure}
\caption{\label{fig:TrackRecoPerf}  (a) The average efficiency to reconstruct primary tracks with a production vertex before the first layer in jets as a function of jet $p_T$. The same sample generation is used for both reconstruction algorithms resulting in correlated features. (b) B-jet efficiency at the baseline 70\% working point of the IP3D algorithm~\cite{ref:IP3D} as a function of truth-jet transverse momentum. The truth jets are reconstructed from generator-level particles in $Z'$ events, using the anti-$k_t$ algorithm~\cite{ref:AntiKt} with $R=0.4$, and are required to be within $|\eta| < 2.5$. 
The performance with the new (TIDE) reconstruction (red squares) is compared with that using the baseline reconstruction from Run-1 (green triangles)~\cite{ref:TIDE}. }
\end{figure}

Following these changes, the tracking efficiency is significantly improved, especially in dense environments where multiple tracks cross the same areas of the detector layers. 
Figure~\ref{fig:DecayProductsEfficiency} shows an improved algorithmic reconstruction efficiency in simulated $\rho\to\pi^+\pi^-$ and $\tau^\pm \to \pi^+\pi^-\pi^\pm\nu_\tau$ events; the efficiency is significantly higher in particular in the high $p_T$ region, where the mother particles are boosted.
Figure~\ref{fig:TrackRecoPerf} shows the improvement in track reconstruction and $b$-jet efficiencies, two quantities that impact physics analyses directly~\cite{ref:TIDE}.

\subsection{\label{sec:bTaggingPerf}Improvements in $b$-jet identification}

\begin{figure}
\begin{subfigure}{.5\linewidth}
\includegraphics[width=\linewidth]{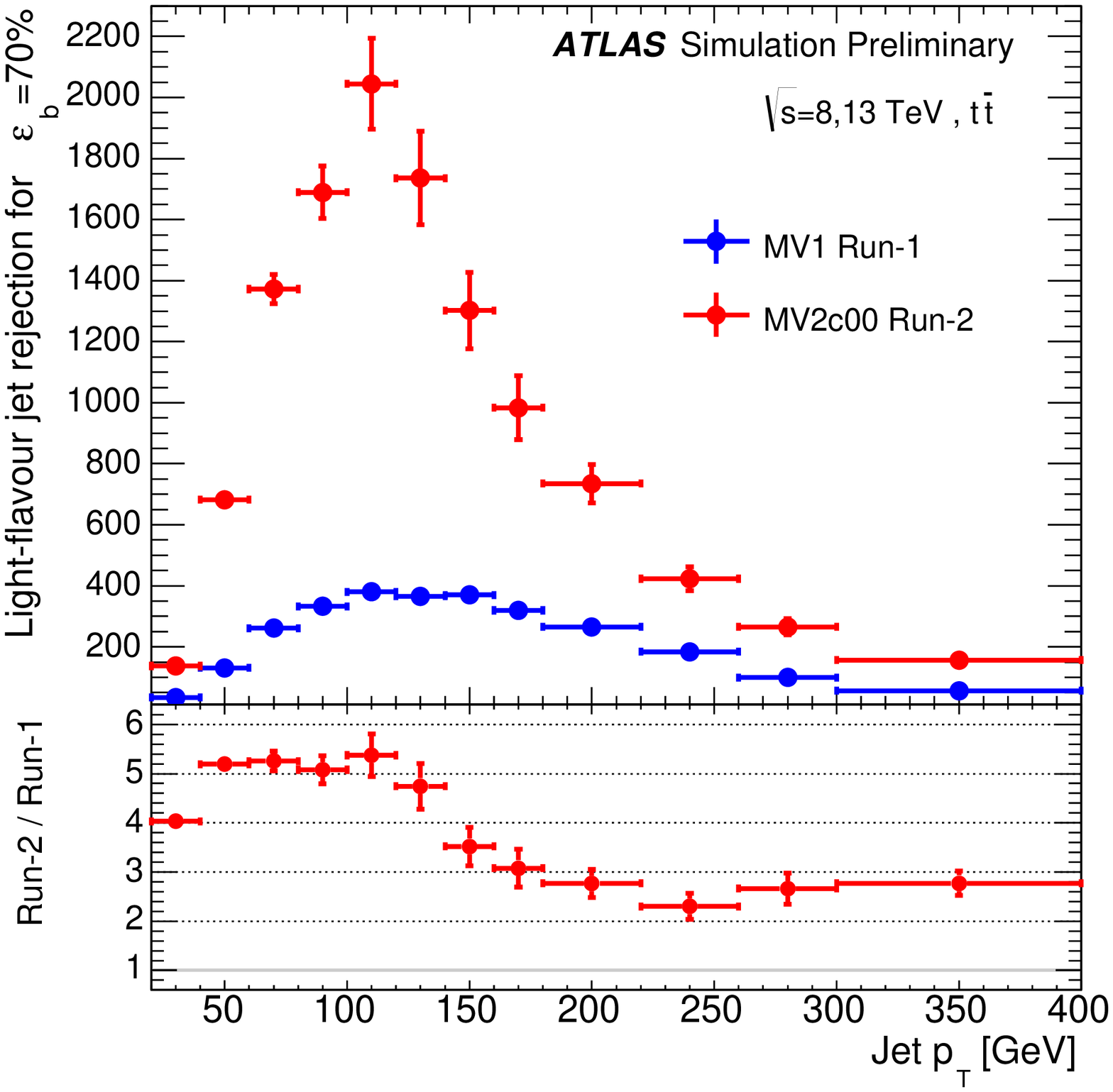}
\caption{\label{fig:bTaggingPerf:a}}
\end{subfigure}
\hspace*{\fill} %
\begin{subfigure}{.5\linewidth}
\includegraphics[width=\linewidth]{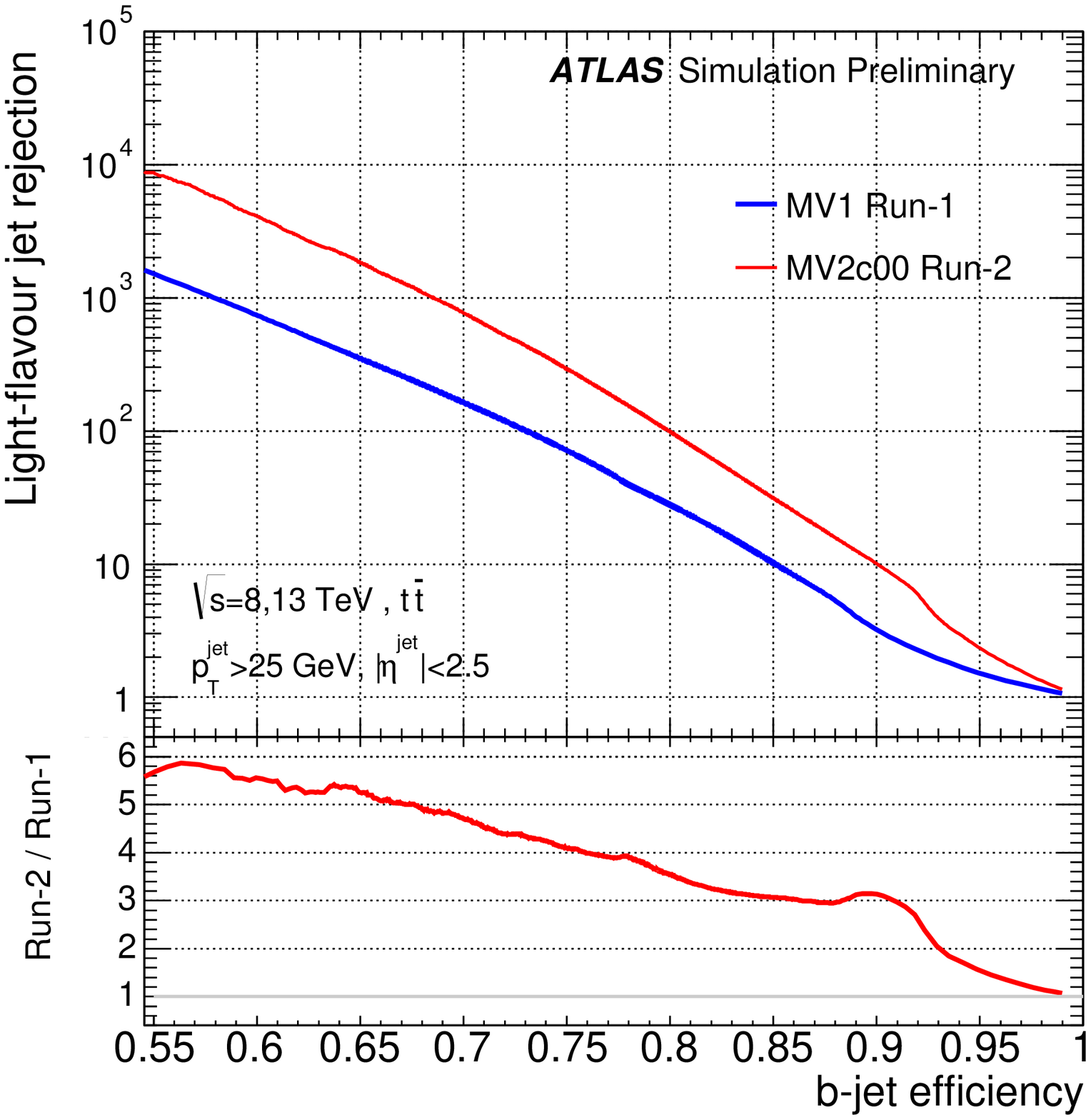}
\caption{\label{fig:bTaggingPerf:b}}
\end{subfigure}
\caption{\label{fig:bTaggingPerf}(a) The light-flavour jet rejection as a function of jet $p_T$ for the MV1 $b$-tagging algorithm using the Run-1 detector and reconstruction software (blue) compared to the MV2c00 b-tagging algorithm using the Run-2 setup (red). In each $p_T$ bin the $b$-tagging cut value has been chosen in such a way to yield a constant b-jet efficiency of 70\%.
(b) The light-flavour jet rejection versus b-jet efficiency for the MV1 b-tagging algorithm using the Run-1 detector and reconstruction software (blue) compared to the MV2c00 b-tagging algorithm using the Run-2 setup (red)~\cite{ref:bTaggingPerformance}.}
\end{figure}

Improvements in tracking have a direct impact on vertex identification. The identification of the primary vertex plays an important role in $b$-jet identification ($b$-tagging), which in turn significantly improves the sensitivity of many analyses. 
For Run-2, in addition to the enhancements to the tracking algorithms, $b$-tagging algorithms were improved. In particular, MV1, the main $b$-tagging algorithm used during Run-1 was improved to obtain MV2. Figure~\ref{fig:bTaggingPerf} shows the effect of these algorithmic improvements on the light-flavour jet rejection as a function of $b$-jet efficiency and jet $p_T$~\cite{ref:bTaggingPerformance}.

\section{\label{sec:Commissioning}Pixel Detector and ID tracking commissioning with early Run-2 data}

In 2015, the ATLAS detector was re-commissioned using cosmic rays and $pp$ collision data. This section presents a few of the many results showing a good performance and understanding of the Pixel Detector and ID tracking. 

\begin{figure}
\begin{subfigure}{.5\linewidth}
\includegraphics[width=\linewidth]{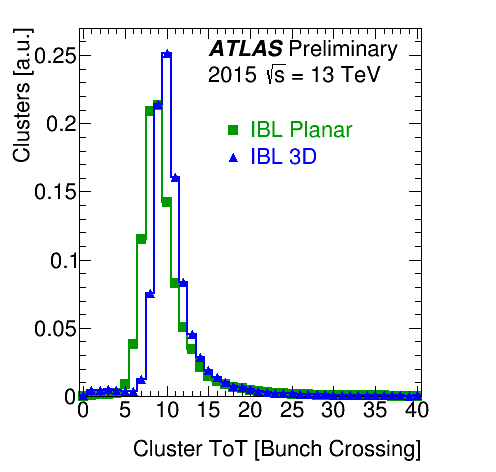} %
\caption{\label{fig:PixelClusterProperties:a}}
\end{subfigure}
\hspace*{\fill} %
\begin{subfigure}{.5\linewidth}
\includegraphics[width=\linewidth]{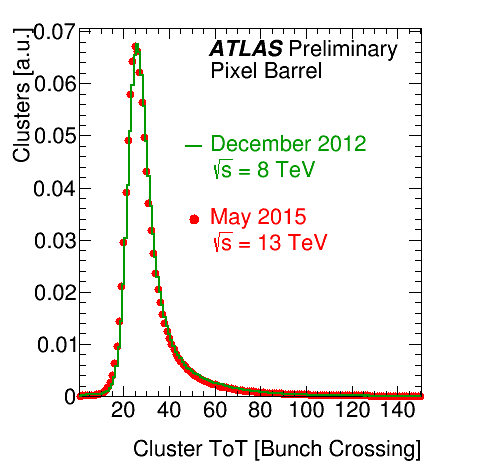} %
\caption{\label{fig:PixelClusterProperties:b}}
\end{subfigure}
\caption{\label{fig:PixelClusterProperties} (a) Cluster on-track time over threshold (ToT) distribution for the two IBL sensor technologies: Planar and 3D sensors. %
Tracks were selected if they had at least 15, 6 and 2 hits in the TRT, SCT and Pixel (including IBL) respectively. Good quality on-track clusters were selected if they are away from the sensor active edge by more than 1.5~mm. %
The different Landau peak position is due to the thickness difference between the two sensors: $0.200$~mm for Planar and $0.230$~mm for 3D sensors. 
(b)  Cluster on-track ToT distributions for the three pixel barrel layers, during the recent 13 TeV run and one of the last $pp$ runs of 2012. The distribution reproduces fairly well the measured one in December 2012. The slight observed shift is under investigation. In both plots, ToT is corrected to normal incidence and the distributions are normalized to unit area~\cite{ref:IblClusterProperties}.
}
\end{figure}

\begin{figure}
\begin{subfigure}{.5\linewidth}
\includegraphics[width=\linewidth]{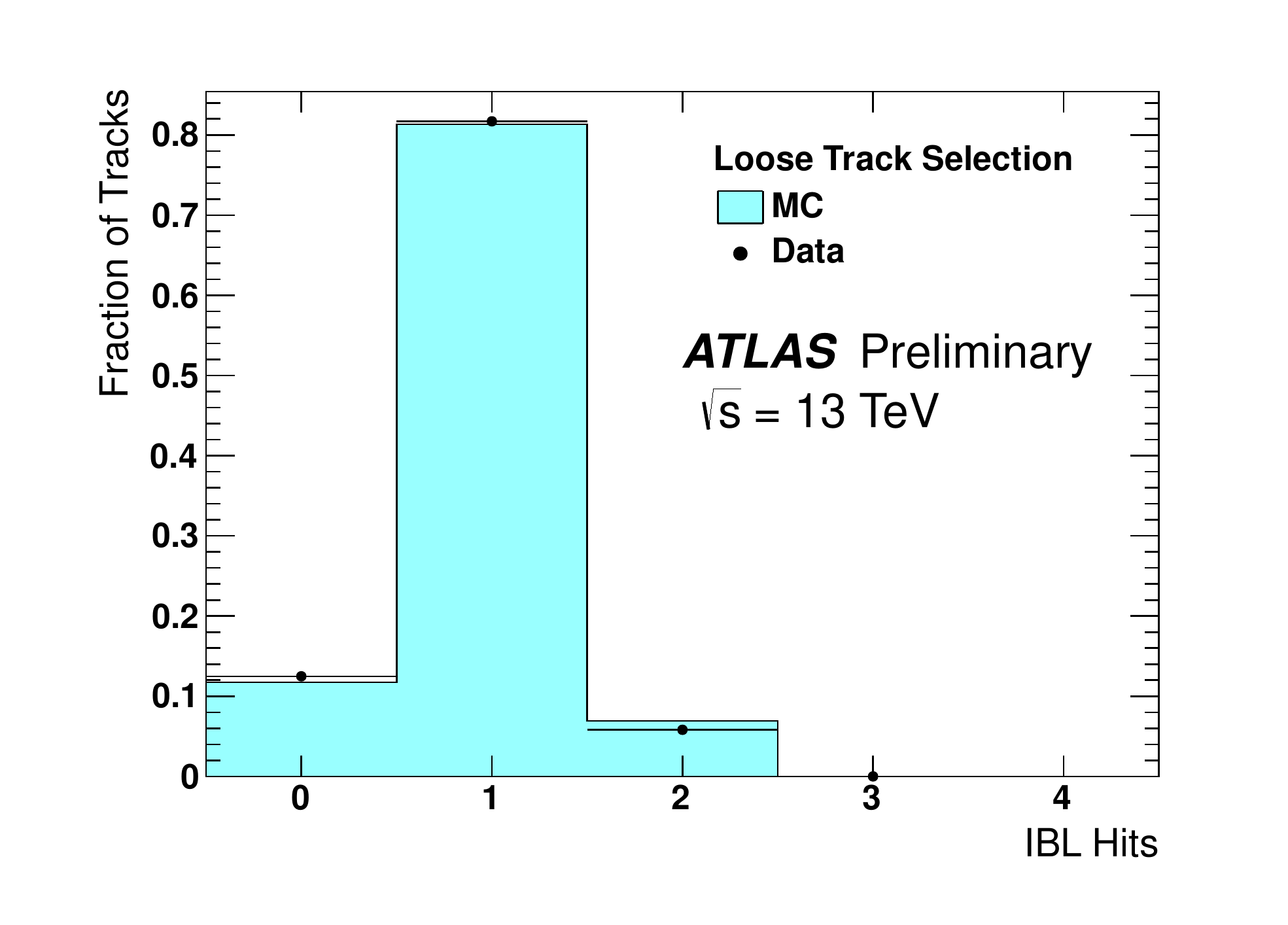}
\caption{\label{fig:TrackReco:a}}
\end{subfigure}
\hspace*{\fill} %
\begin{subfigure}{.5\linewidth}
\includegraphics[width=\linewidth]{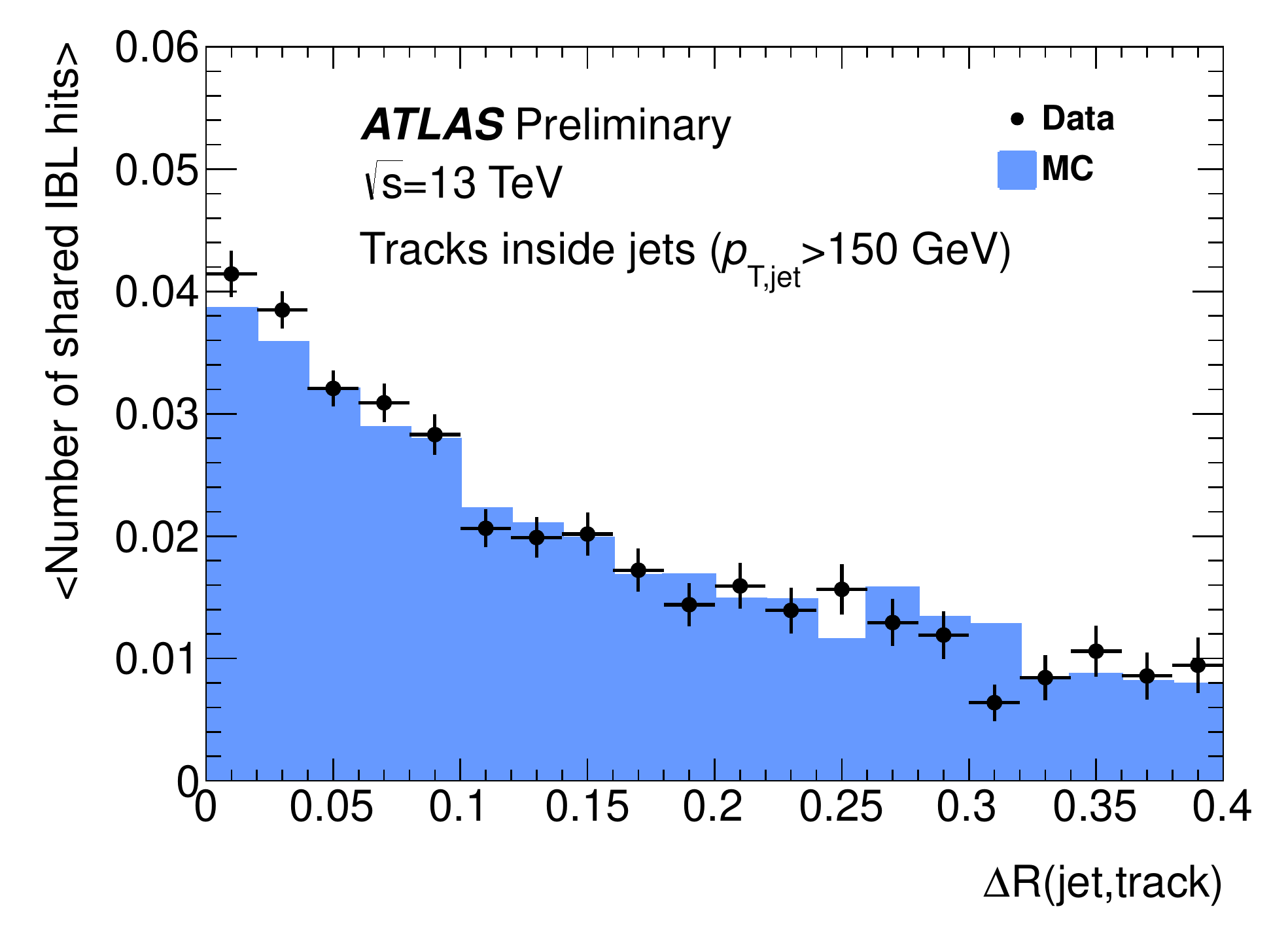}
\caption{\label{fig:TrackReco:b}}
\end{subfigure}
\caption{\label{fig:TrackReco}  (a) Comparison of the number of IBL hits in data and simulation (PYTHIA 8 A2:MSTW2008LO) for the loose track selection~\cite{ref:TrackRecoPerfInID}. The distribution is normalized to one. 
(b) The number of shared IBL hits on track as a function of $\Delta R$(jet,track) in data and simulation. 
Shared hits are used by more than one track and are not identified as split. Only tracks inside jets with a $p_T>150$~GeV are selected~\cite{ref:TrackRecoPerfInID}.}
\end{figure}

Figure~\ref{fig:PixelClusterProperties} shows the performance of the Pixel Detector in early 13~TeV data. The cluster on-track time over threshold (ToT) is shown for the IBL as well as for the Run-1 Pixel Detector, together with a comparison with Run-1~\cite{ref:IblClusterProperties}. These match the expectations.
Figure~\ref{fig:TrackReco} shows a comparison between data and simulation of the distributions of the number of IBL hits and of the number of shared IBL hits on track as a function of $\Delta R({\rm jet, track})$. A good agreement is seen in these and other distributions of quantities relevant to track reconstruction shown in~\cite{ref:TrackRecoPerfInID}.

\begin{figure}
\centering
\includegraphics[width=.5\linewidth]{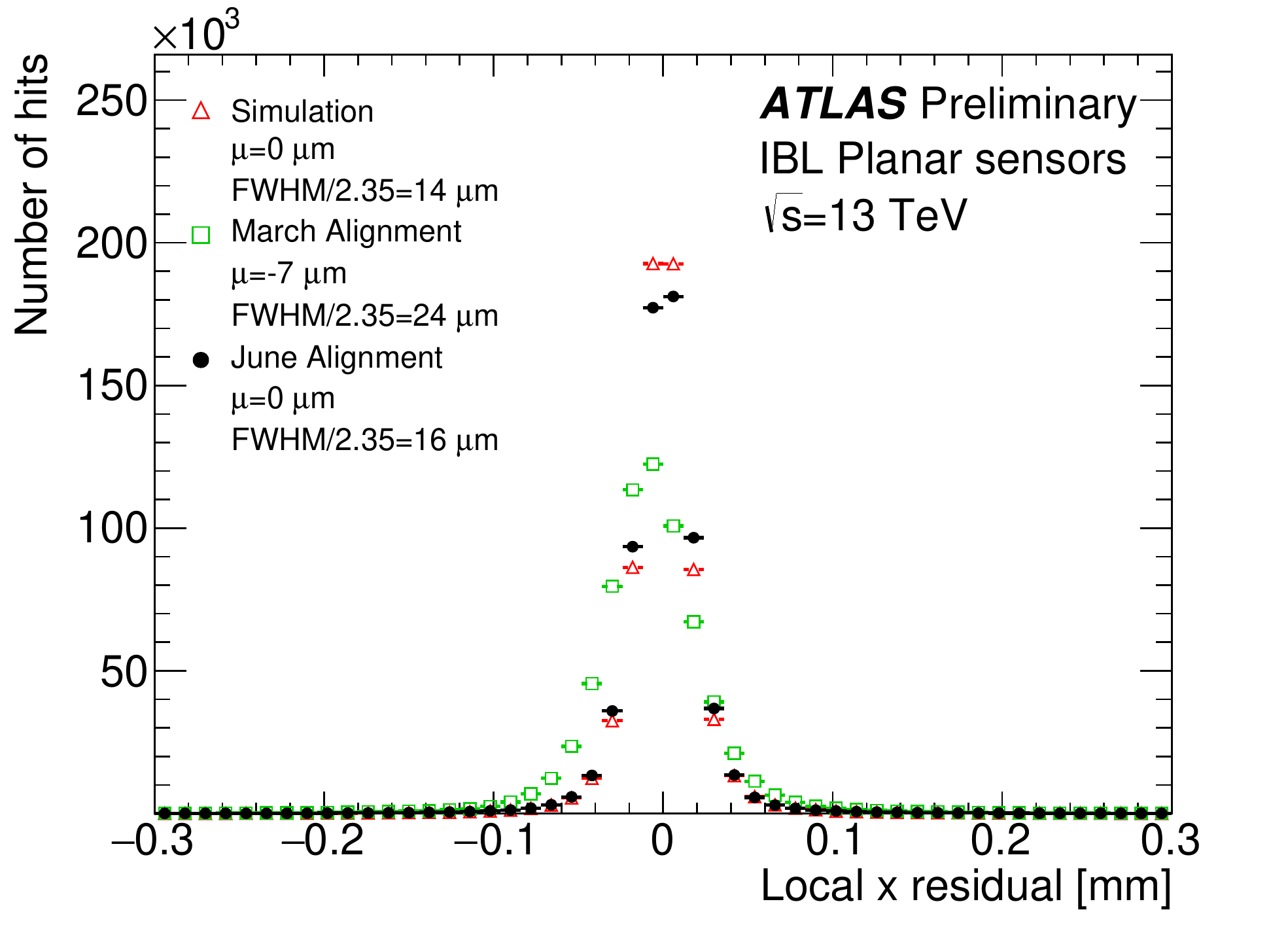}
\caption{\label{fig:ResXDistributionIblPlanar}The local-$x$ residual distribution for the IBL planar sensors for the 13~TeV collision data sample reconstructed with the June alignment (black) and March alignment (green) as well as observed in the perfectly aligned simulation (red). The distributions are integrated over all hits assigned to tracks in the respective IBL regions. The parameter $\mu$ represents the mean of the distributions. The distributions have been normalized to the same number of entries~\cite{ref:InitialAlignmentID}.}
\end{figure}

\begin{figure}
\begin{subfigure}{.5\linewidth}
\includegraphics[width=\linewidth]{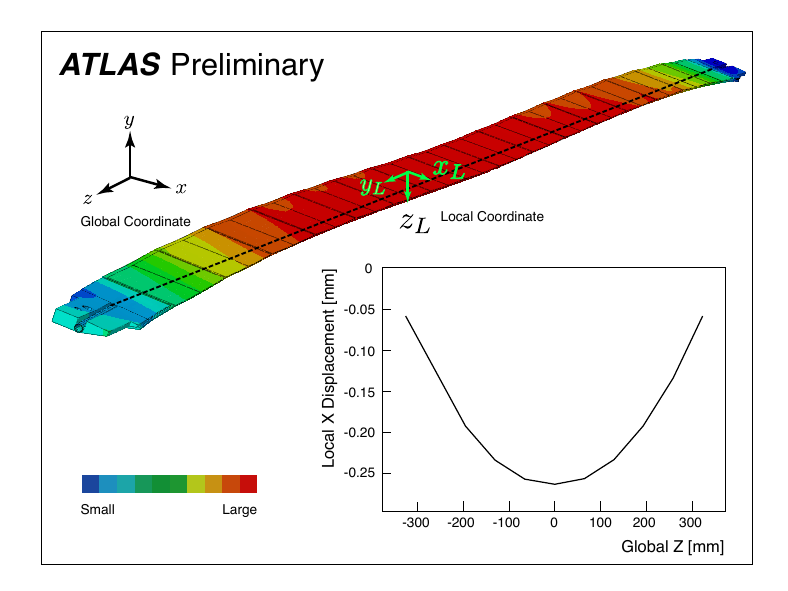}
\caption{\label{fig:IblMechanicalStability:a}}
\end{subfigure}
\hspace*{\fill} %
\begin{subfigure}{.5\linewidth}
\includegraphics[width=\linewidth]{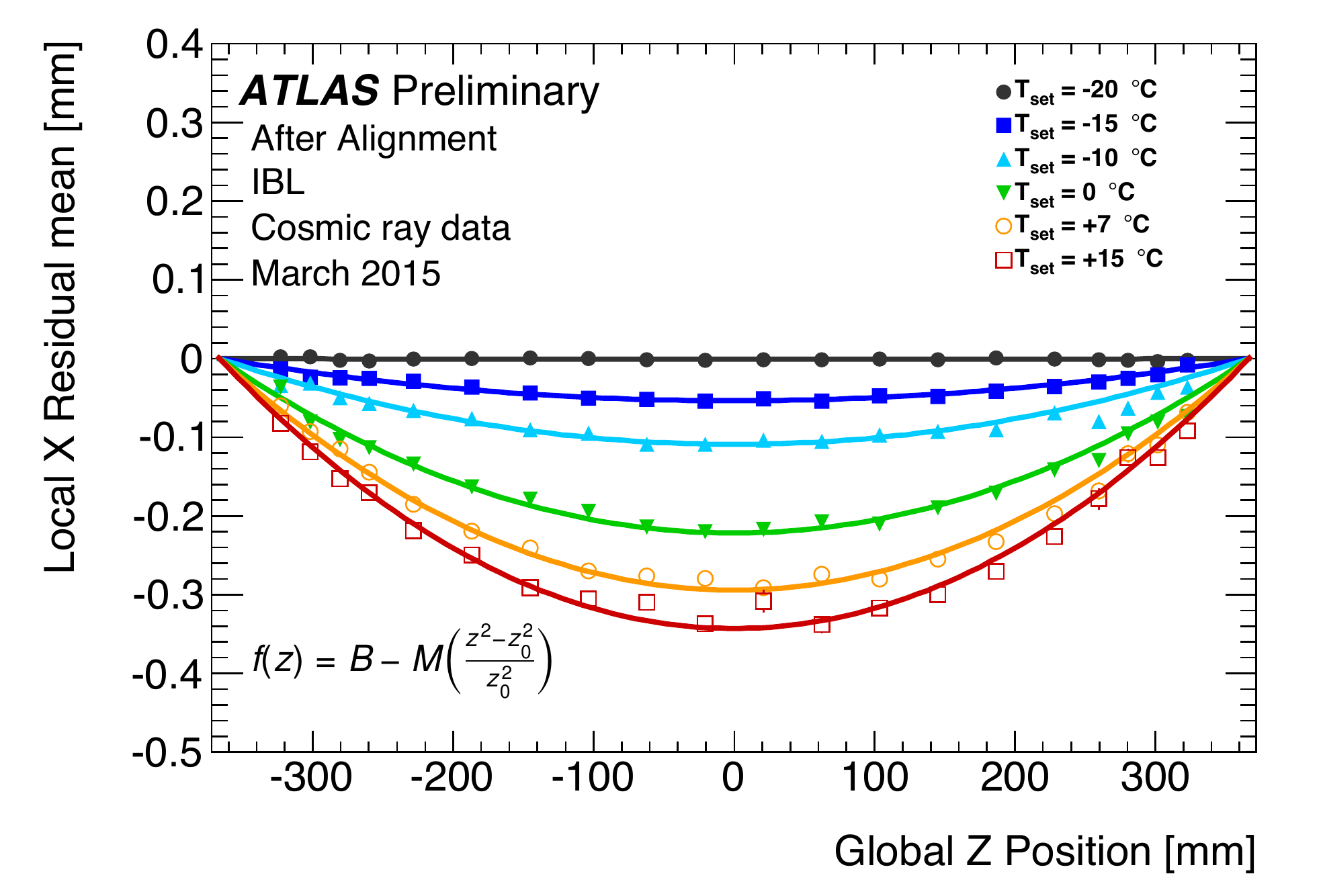}
\caption{\label{fig:IblMechanicalStability:b}}
\end{subfigure}
\caption{\label{fig:IblMechanicalStability} (a) Visualization of the distorted stave. The size of the distortion is magnified for visualization. The color represents the magnitude of the displacement. The right bottom graph shows the relative displacement size in local-$x$ direction ($x_L$) as a function of the global $z$-position at the face plate surface of the stave.
(b) The track-to-hit residual mean in the local-$x$ direction. The residual mean is averaged over all hits of modules at the same global-$z$ position. The alignment corrections derived at -$20^\circ$C are applied to the local positions in the module frames. For local-$x$, each data set is fitted to a parabola which is constrained to match to the baseline $B=0$ at $z=\pm z_0 = \pm 366.5$~mm~\cite{ref:iblMechanicalStability}.}
\end{figure}

\begin{figure}
\centering
\includegraphics[width=.5\linewidth]{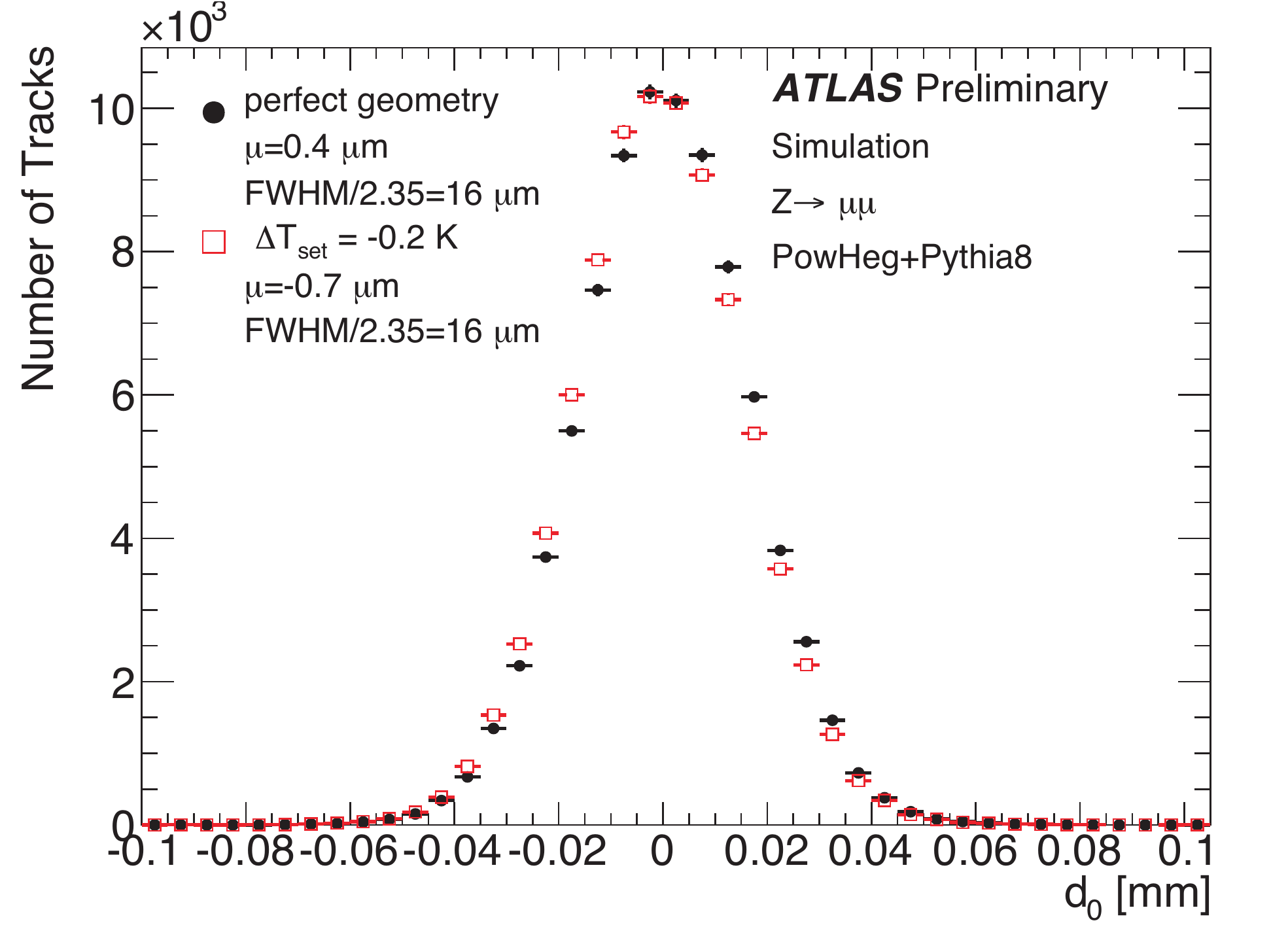}
\caption{\label{fig:ImpactParameterDistortion} Distribution of the transverse impact parameter $d_0$ of charged tracks with respect to the beam-spot from $Z\to\mu^+\mu^-$ events simulated in $\sqrt{s}=13$~TeV $pp$ collisions. The solid circle shows the nominal geometry and the open square shows the distorted geometry corresponding to a temperature variation of the IBL at -0.2 K ($\sim2\mu$m of displacement at the center of the stave)~\cite{ref:iblMechanicalStability}.}
\end{figure}

Figure~\ref{fig:ResXDistributionIblPlanar} shows the consecutive improvements in ID alignment~\cite{ref:AlignmentPerfID,ref:InitialAlignmentID} using the local-$x$ distribution for IBL as the various alignment steps are performed. During the alignment of the ID, a mechanical distortion of the IBL was observed. This distortion, shown in figure~\ref{fig:IblMechanicalStability}, is caused by a difference in the coefficients of thermal expansion of the IBL stave components~\cite{ref:iblMechanicalStability}. The magnitude of the distortion is found to depend linearly on the operating temperature of the IBL, with a gradient of $\sim10 \mu$m/K. The expected bias to the transverse impact parameter ($d_0$) of charged tracks under a temperature variation of 0.2~K is evaluated to be $\sim1 \mu$m using $Z\to\mu^+\mu^-$ events from a Monte Carlo simulation of $pp$ collisions at $\sqrt{s}=13$~TeV, as shown in figure~\ref{fig:ImpactParameterDistortion}. This bias is small in comparison to the expected $d_0$ resolution of $O(10~\mu{\rm m})$. The temperature of the IBL detector is constantly monitored and this information is taken into account during the detector alignment~\cite{ref:iblMechanicalStability}.

\begin{figure}
\begin{subfigure}{.5\linewidth}
\includegraphics[width=\linewidth]{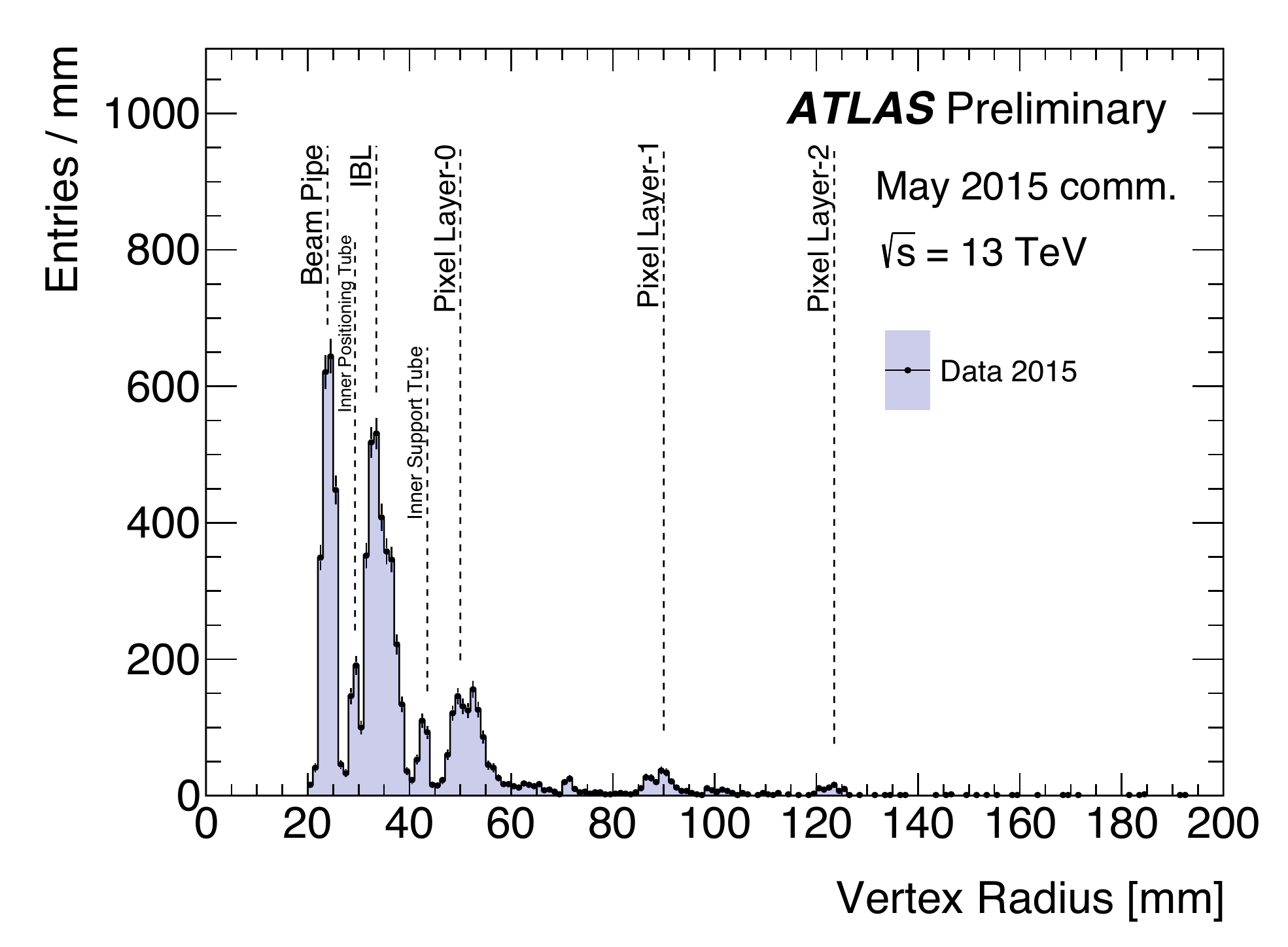}
\caption{\label{fig:HadInt:a}}
\end{subfigure}
\hspace*{\fill} %
\begin{subfigure}{.5\linewidth}
\includegraphics[width=\linewidth]{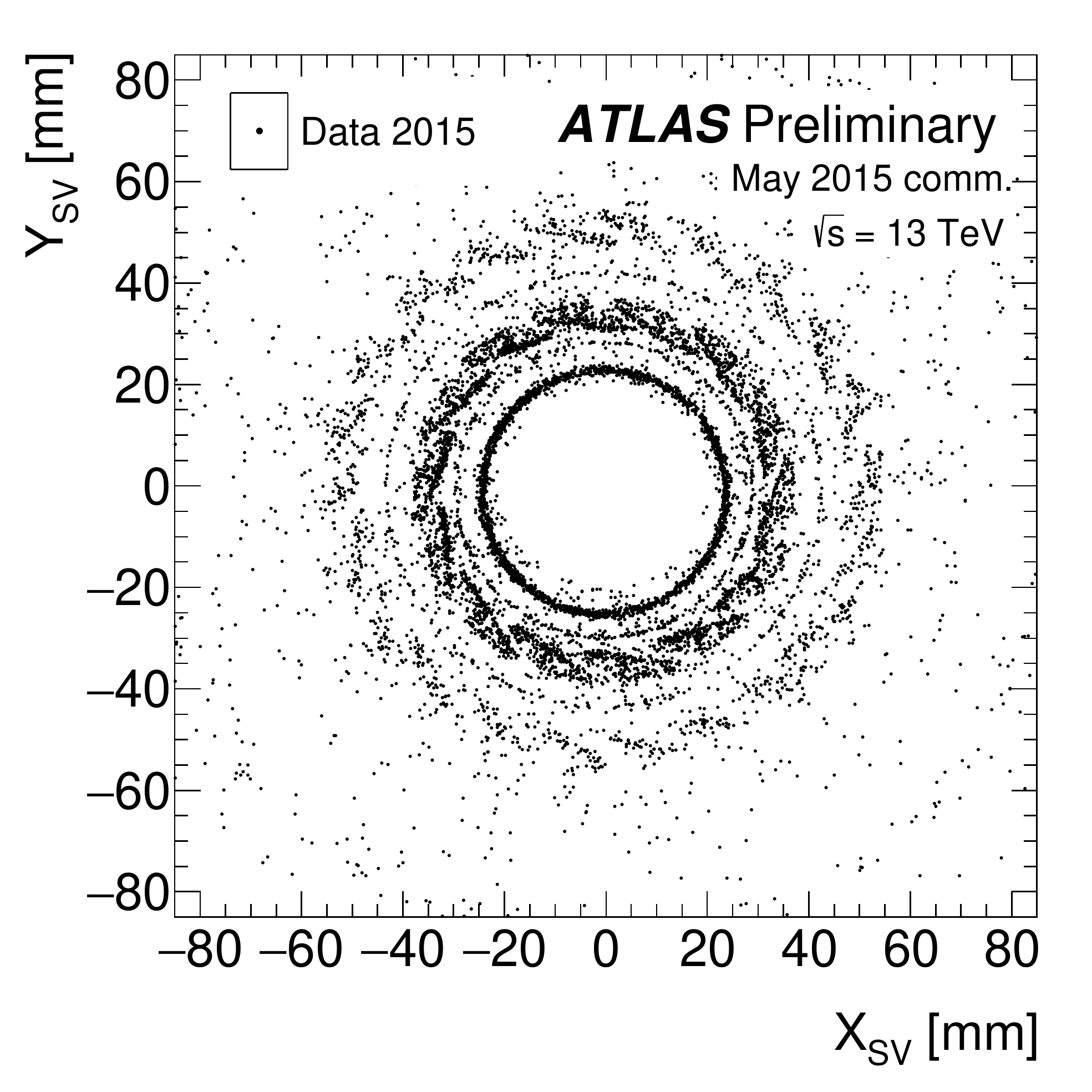}
\caption{\label{fig:HadInt:b}}
\end{subfigure}
\caption{\label{fig:HadInt}  (a) The radial vertex position distribution and (b) the vertex position distribution in the $xy$-plane for hadronic interaction candidates reconstructed from multiple tracks with $>5\sigma$ transverse impact parameter significance. The events were collected in May 2015 during the commissioning of the Run-2 ATLAS detector. The distribution exhibits peaks consistent with hadronic interactions occurring within the new ATLAS beam pipe and the Insertable B-Layer (IBL) in addition to the original three Pixel layers. The hadronic interaction candidates are required to be reconstructed within $r>20$~mm and $|z|<300$~mm. Due to impact parameter cuts, the efficiency depends on the radius. No attempt has been made to correct for this~\cite{ref:IDTR-2015-003}.}
\end{figure}

A good understanding of the material budget of the ATLAS Inner Detector is crucial for physics analyses. For this purpose a careful mapping of the ID material is performed. Figure~\ref{fig:HadInt} shows preliminary results of one of such mappings using hadronic interactions in early {\it quiet beam} data collected in May 2015. One clearly sees the ATLAS beam pipe and the 4 Pixel Detector layers~\cite{ref:IDTR-2015-003}.

\begin{figure}
\begin{subfigure}{.5\linewidth}
\includegraphics[width=\linewidth]{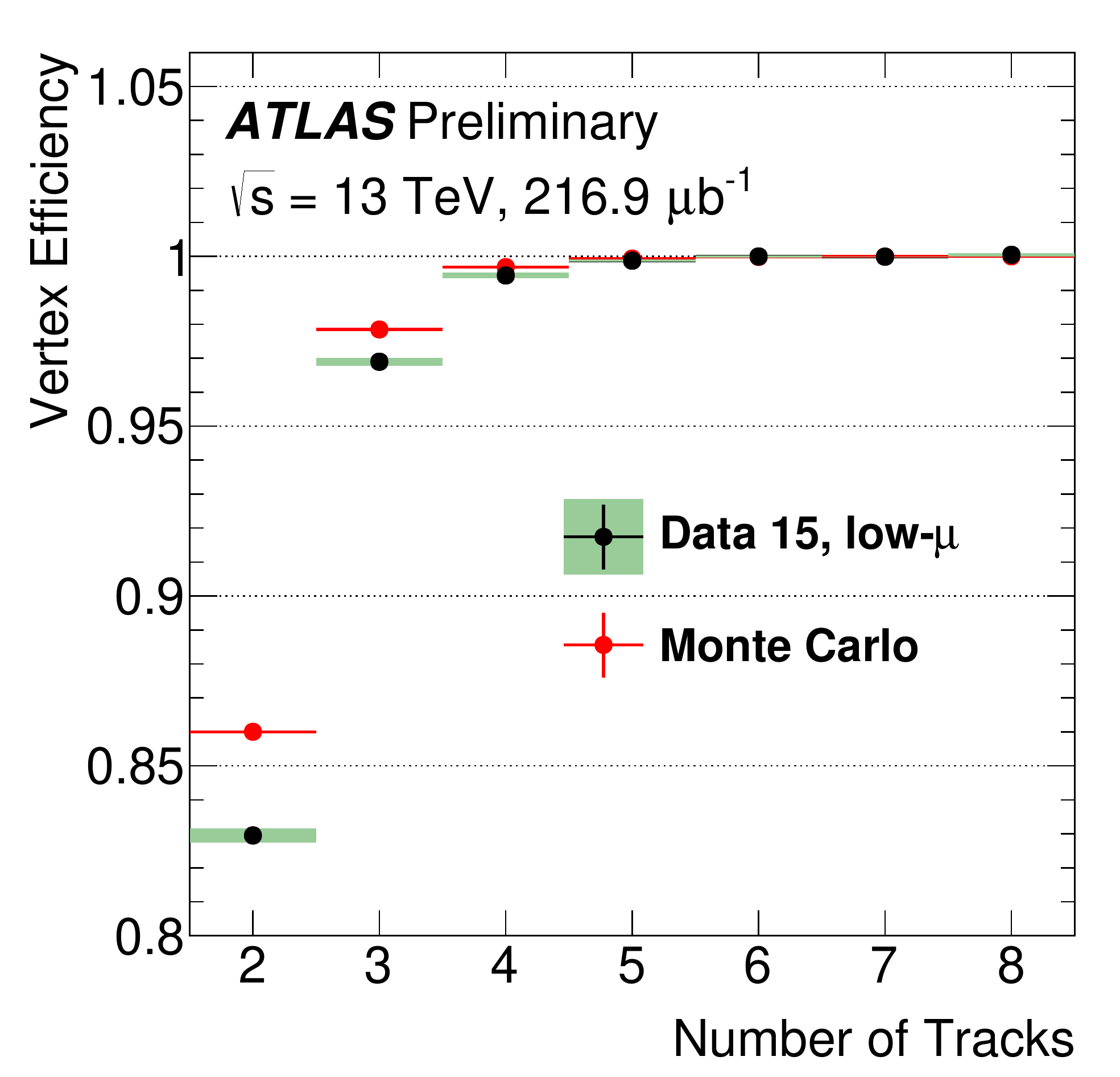}
\caption{\label{fig:VertexReco:a}}
\end{subfigure}
\hspace*{\fill} %
\begin{subfigure}{.5\linewidth}
\includegraphics[width=\linewidth]{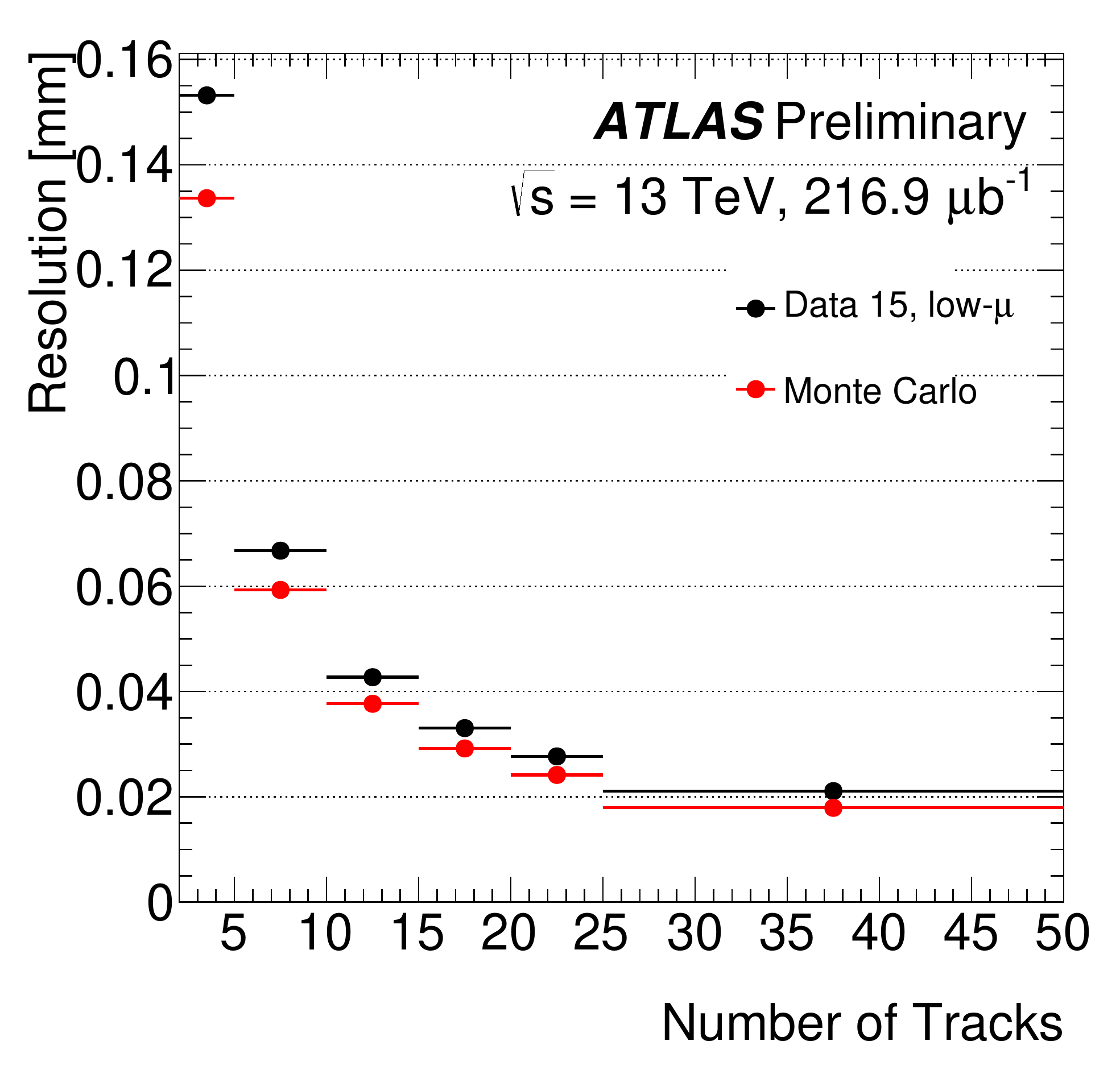}
\caption{\label{fig:VertexReco:b}}
\end{subfigure}
\caption{\label{fig:VertexReco}  (a) Vertex reconstruction efficiency as a function of the number of tracks and (b) primary vertex resolution (reconstruction resolution $\sigma_x$ corrected by a scale-factor) as a function of the average number of tracks in the two vertices used in the Split-Vertex method.
The distributions are shown for low-$\mu$ (low pile-up) data compared to Monte Carlo simulation (Pythia 8 A2:MSTW2008LO). These measurements use a subset of the low-$\mu$ dataset corresponding to an integrated luminosity of $216.9$~$\mu\rm{b}^{-1}$~\cite{ref:VertexRecoPerformance}.}
\end{figure}

The primary goal of the ID tracking is to provide good vertex reconstruction efficiency and (primary) vertex resolution. Figure~\ref{fig:VertexReco} shows these 2 distributions as a function of the number of tracks for low-$\mu$ (low pile-up) data and compares them with simulation. Additional details can be found in~\cite{ref:VertexRecoPerformance}. 

\begin{figure}
\centering
\includegraphics[width=.5\linewidth]{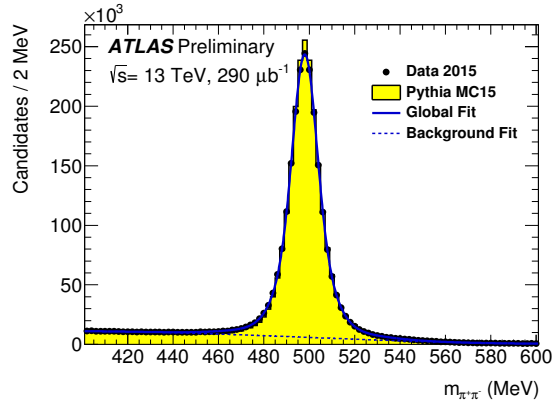}
\caption{\label{fig:KShort}Comparison of measured and predicted $K_S$ invariant-mass distributions in 13 TeV samples. The points are data, while the histograms show the Monte Carlo sample with signal and background components separately normalised to the data. The solid line is the line-shape function fitted to data, the dashed line shows the component of the fitted function describing the combinatorial background~\cite{ref:IDTR-2015-006}.}
\end{figure}

As an example of physical observables who rely on good tracking performance, Figure~\ref{fig:KShort} shows the distribution of measured $K_S$ invariant mass in early Run-2 data and the good agreement with the prediction form Monte Carlo simulation~\cite{ref:IDTR-2015-006}.

\clearpage

\section{Summary and Outlook}

For Run-2, the ATLAS Pixel Detector was upgraded with a 4th layer, the IBL, and a new beam monitor, DBM. The reconstruction of charged particle tracks -- in particular the impact parameter resolution -- was significantly improved in Run-2, both due to the new detector layer and to algorithmic and software improvements, yielding a fourfold decrease of per-event processing time. Furthermore, tracking in dense environments was improved and provides an increased efficiency. These improvements provide a strong basis for physics analyses using Run-2 ATLAS data at $\sqrt{s} = 13$~TeV.

\end{document}